\documentclass[twocolumn]{aastex631}

\usepackage{xspace}
\usepackage{amsmath}

\usepackage[normalem]{ulem}

\begin{document}
%% Front Matter
\title{A Unified, Physical Framework for Mean Motion Resonances}
\author[0000-0002-9908-8705]{Daniel Tamayo}
\affiliation{Department of Physics, Harvey Mudd College, 301 Platt Blvd., Claremont, CA, 91711, USA}
\author[0000-0002-1032-0783]{Samuel Hadden}
\affiliation{Canadian Institute for
Theoretical Astrophysics, 60 St George St Toronto, ON M5S 3H8, Canada}
\begin{abstract}
The traditional approach to analyzing mean motion resonances is through canonical perturbation theory.
While this is a powerful method, its generality leads to complicated combinations of variables that are challenging to interpret and require looking up numerical coefficients particular to every different resonance. 
In this paper we develop simpler scaling relations in the limit where orbits are closely spaced (period ratios $\lesssim 2$) and interplanetary interactions can be approximated by only considering the close-approaches each time the inner planet overtakes the outer at conjunction.
We develop geometric arguments for several powerful results: (i) that $p$:$p-q$ MMRs of the same order $q$ are all rescaled versions of one another (ii) that the general case of two massive planets on closely spaced, eccentric, co-planar orbits can be approximately mapped onto the much simpler case of an eccentric test particle perturbed by a massive planet on a co-planar circular orbit and (iii) that while the effects of consecutive conjunctions add up coherently for first-order ($p$:$p-1$) MMRs, they partially cancel for $p$:$p-q$ MMRs with order $q>1$, providing a physical explanation for why these higher order MMRs are weaker and can often be ignored. 
Finally, we provide simple expressions for the widths of MMRs and their associated oscillation frequencies that are universal to all closely spaced MMRs of a given order $q$, in the pendulum approximation.

\end{abstract}

\section{Introduction}

Mean motion resonances (MMRs) between a pair of planets' orbital periods play a seemingly contradictory role in planetary dynamics.
On the one hand, they can be powerful stabilizing forces.
For example, the crossing orbits of Neptune and Pluto suggest that the two bodies should eventually meet and either collide or scatter; however, the 3:2 resonance between their orbital periods sets up a repeating dance that ensures that Neptune and Pluto never meet at a point of intersection.
On the other hand, resonances are also at the root of chaos and instabilities in planetary dynamics.
For example, the Kirkwood gaps in the asteroid belt correspond to locations with integer period ratios with Jupiter, and are regions that have been dramatically depleted of asteroids.

This dichotomous behavior can be understood by analogy between the orbital dynamics near resonance and those of a pendulum.
A pendulum hanging straight down is stable to additional perturbations, and will simply oscillate around the stable equilibrium point; this situation corresponds to Neptune and Pluto's orbital configuration, which is stable to small tugs from other planets.
On the other hand, a rigid pendulum standing upside down is in an unstable configuration. 
The slightest perturbation is enough to send it falling one way or the other, and when it returns, it once again becomes sensitive to any additional unmodeled details in the dynamics that might send it ``over the top" or back the way it came.
This erratic behavior and sensitivity to small additional perturbations is the hallmark of chaos, and opens up the possibility of instabilities in conservative systems.

This pendulum analogy for MMRs has long been discussed \citep[e.g.,][]{Goldreich65}, and is traditionally reached through the celestial mechanics method of isolating resonant harmonics from the ``disturbing function" expansion of the interplanetary gravitational potential.

This mathematical approach is powerful, but provides little physical intuition.
Most dynamicists would be hard-pressed to answer many seemingly simple questions.
What is the physical mechanism for period ratio oscillations near MMRs?
Does one need to treat each $p$:$p-q$ MMR separately, or is there some underlying relationship connecting them?
Why is it that at low eccentricities, the strongest MMRs have integer ratios of the form $p$:$p-1$?
In the general case where both planets are massive and on eccentric orbits, what is the meaning of the linear combination (specific to each particular MMR) of masses and eccentricities that affects the dynamics, and why is there a separate one that is approximately conserved?

A setting in which the dynamics of the planetary three-body problem become simpler, and provides significant physical insight, is the so-called ``Hill limit", where the two planetary orbits are closely spaced.
In this limit, the gravitational interaction becomes dominated by the close approaches at conjunction each time the inner object overtakes the outer and the distance between the two bodies is minimized.
The dynamics of these close encounters have been carefully analyzed \citep[e.g.,][]{Hill1878, Henon66, Henon86}, and several authors have additionally analyzed the long-term effects of repeated encounters \citep[e.g.,][]{Duncan89, Namouni96}.
As we review below, this model provides a clear physical picture of how exchange of orbital energy leads to variations in orbital period.
However, such works largely focus on the general dynamics rather than on the specific case of resonant initial conditions. 

\cite{Hadden19} realized that the fact that the celestial mechanics approach at arbitrary separations must match up with the Hill limit results at close separations implies an important connection between these two problems.
This helped him to generalize a powerful result for the dynamics of co-planar planets near a first-order $p$:$p-1$ MMR \citep{Sessin84} to arbitrary $p:p-q$ MMRs: that the problem
can be reduced from four degrees of freedom (for each planet's semimajor axis and eccentricity) down to one, whose dynamics are similar to those of a pendulum.
These results are derived using formal methods of canonical perturbation theory, which lead to precise but sometimes cumbersome expressions that provide limited physical intuition. 

In this paper, we explore the above results instead using simple scaling arguments, building from a simple model by \cite{Peale76, Greenberg77, Peale86, Murray99} for first-order MMRs (Sec.\:\ref{sec:qualitative}).
The primary new result in this paper is that higher order MMRs are weaker than first-order MMRs due to the canceling effects of multiple conjunctions per cycle (Sec.\:\ref{sec:higher}).
In addition, we provide (to our knowledge) a novel scaling argument for the widths and oscillation frequencies inside first-order MMRs (Sec.\:\ref{sec:scaling}).
More broadly, we aim to develop a physical understanding of the powerful result that (a) all MMRs of a given order can be put on the same footing \citep[][Sec. \:\ref{sec:normalizedMMRs}]{Hadden18} and (b) that the general case of MMRs between two massive planets on eccentric, planar orbits can be approximately mapped to the much simpler PCR3BP for closely spaced orbits \citep[][Sec.\:\ref{sec:general}]{Hadden19}.
This provides a clear physical meaning for the important combinations of orbital parameters in the Hill limit, and can serve as an easily understood vocabulary with which to understand and discuss canonical variables for MMRs at wider separations.
A summary of the main results is presented up front in Sec.\:\ref{sec:summary}.

\section{A Qualitative Model for First-Order MMRs in the PCR3BP} \label{sec:qualitative}

\begin{figure*}
    \centering
    \resizebox{0.99\textwidth}{!}{\includegraphics{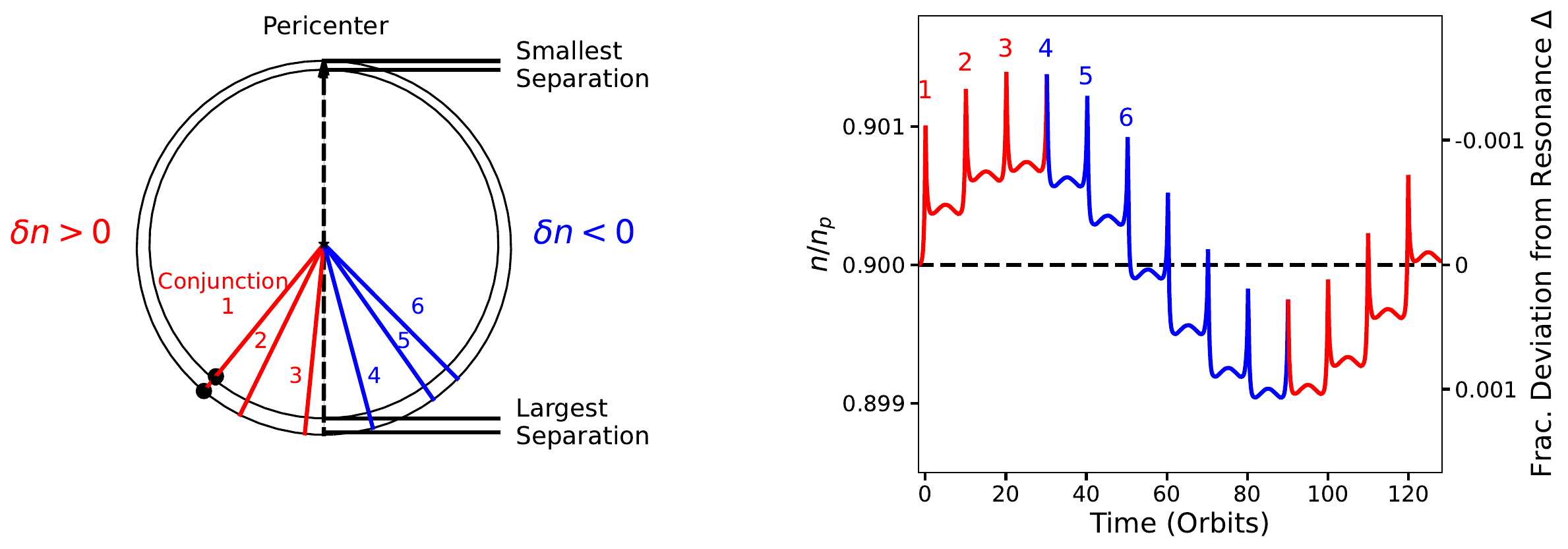}}
    \caption{Geometry of resonance: Massive inner planet is on a fixed circular orbit. Outer test particle is on an eccentric orbit with the pericenter always pointing upward (left panel). Both bodies orbit counterclockwise. The test particle is initialized with its mean motion at a 9:10 ratio with the inner planet, with the first conjunction occurring where the planets are plotted in the left panel. Succcessive conjunctions are numbered in the left panel with lines pointing to their locations. Conjunctions to the left of the line of apses (dashed line) increase the test particles mean motion (red), while conjunctions to the right decrease it (blue)---see text for details. A full cycle for the evolution of the ratio of mean motions is plotted in the right panel vs time, labeling the first 6 conjunctions that are plotted in the left panel. The dynamics are analogous to the oscillations of a pendulum as described in the text. The rightmost axis plots $\Delta$ (Eq.\:\ref{eq:Delta}), the fractional deviation from resonance. 
    \label{fig:qualitative}}
\end{figure*}

%We begin by restricting ourselves to first-order ($j$:$j-1$) MMRs, where each cycle, the inner planet does one more orbit than its outer neighbor, resulting in a single conjunction.
%After analyzing the effects of these individual conjunctions we will apply this understanding to higher order MMRs in Sec.\:\ref{sec:higher}.

The key physical insight in the Hill limit is that the gravitational interaction between the secondaries becomes dominated by the close approaches at conjunction \citep{Hill1878, Henon86}.
In this paper, we will restrict ourselves to the typical regime of interest where the interplanetary forces nevertheless remain much smaller than those from the central star and can be treated as a perturbation.
In fact, it is valuable to begin by neglecting the interaction altogether.

Consider the case of a 10:9 MMR, with conjunctions occurring at the arbitrary longitude labeled Conjunction 1 in Fig.\:\ref{fig:qualitative}, with both planets orbiting counterclockwise.
If the planets don't interact, then the resonance condition implies that the next conjunction will occur after 10 orbits of the inner planet and 9 orbits of our test particle, at exactly the same location.
However, imagine now that the test particle was on a shorter-period orbit so that it moved slightly faster.
Then, after the same 10 orbits of our inner planet, the test particle would now find itself further ahead, so the next conjunction would happen further along the orbit.
Conjunctions would thus steadily advance along the orbit, while the  orbital periods and thus their period ratio would remain constant. 
In this non-interacting limit, MMRs are then infinitely ``thin": the only periodic solutions whose configurations repeat are those with exact integer ratios between their orbital periods.

However, if we turn on the interplanetary interactions by giving the inner planet mass, MMRs acquire finite ``widths", in the sense that period ratios initialized within a small range of the integer ratio will now oscillate about that resonant value. 
A qualitative argument for this behavior is given by \cite{Peale76, Greenberg77, Peale86, Murray99}.
We first recapitulate their arguments, and in Sec.\:\ref{sec:qualitative} derive the relevant scalings through simple arguments.
For simplicity of notation in this discussion of the PCR3BP, we will denote the orbital elements of the unperturbed massive planet with the subscript $p$, and omit subscripts for those of the test particle on the outer eccentric orbit.
The integrations shown in Fig.\:\ref{fig:qualitative} assume an Earth-mass inner planet around a solar-mass star, an initial eccentricity for the outer orbit that is 30\% of the orbit-crossing value, and an initial conjunction occurring $2\pi/3$ radians away from pericenter.

Because the semimajor axis (or equivalently the mean motion through Kepler's third law) depends only on the orbital energy in the Kepler problem, we specifically examine the energy exchange at conjunction when the planets interact significantly.
As the faster-moving inner planet comes from behind to overtake, it gravitationally pulls the test particle backward along its orbital path, removing energy $E$ from the latter's orbit at a rate 
\begin{equation}
\dot{E} = {\bf F}\cdot{\bf v}, \label{eq:Edot}
\end{equation}
where we will denote vectors with boldface and time derivatives with overdots, and ${\bf v}$ is the test particle's velocity, while ${\bf F}$ the gravitational force acting on it.
Conversely, after conjunction, the planet is now ahead of the test particle and because ${\bf v}$ and ${\bf F}$ now point in the same direction, $\dot{E}$ is positive and the test particle gains energy.

While these effects partially cancel out, there is in general a net exchange of energy for an eccentric test particle orbit, due to the fact that the pre and post-conjunction interactions are asymmetric.
Because the separation between the orbits is increasing from pericenter (top of left panel of Fig.\:\ref{fig:qualitative}) to apocenter (bottom), conjunction 1 has a loss of energy pre-conjunction (at closer separations) that is larger than the gain of energy post-conjunction (at larger separation).
The test particle therefore suffers a net loss of energy, causing it to fall deeper in the star's potential well, i.e., to a smaller semimajor axis.
This leads to the counterintuive result for Keplerian orbits that a loss of energy causes a body to orbit {\it faster} (i.e., the kick to the mean motion from the conjunction $\delta n > 0$). 
This will be the case for any conjunction occuring to the left of the vertical dashed line marking the line of apses, while conjunctions to the right (in blue) will result in a net gain of energy and cause the test particle to orbit {\it slower}. 

We can now predict the qualitative orbital evolution in the case of Fig.\:\ref{fig:qualitative}.
Initially, the test particle's mean motion is exactly at the resonant value for the $10:9$ MMR ($n/n_p = 9/10$ in the right panel).
However, because the first conjunction occurs to the left of the line of apses (marked 1 on both panels), the test particle speeds up (the planet's $n_p$ remains constant, so $n/n_p$ rises in the right panel).
Since the test particle is now moving faster than the resonant value, conjunction 2 happens further along the orbit the next time they meet (labeled 2 in both panels), which acts to further speed up the test particle.
Conjunction 3 occurs almost at apocenter, and is thus approximately symmetric pre and post encounter, resulting in little change to the outer test particle's mean motion $n$.
Nevertheless, by this point $n$ has accumulated a significant excess above the resonant value (dashed black line in the right panel), so the conjunction location continues marching counter-clockwise.

Indeed, the further the ratio of mean motions is from the resonant value, the more the conjunction location shifts from one conjunction to the next.
It is therefore convenient to define a relative deviation from resonance $\Delta$, plotted along the right axis of the right panel in Fig.\:\ref{fig:qualitative} \citep{Lithwick12},
\begin{equation}
\Delta \equiv \frac{\frac{n_1}{n_2} - \Bigg(\frac{n_1}{n_2}\Bigg)_\text{res}}{\Bigg(\frac{n_1}{n_2}\Bigg)_\text{res}} = \frac{\frac{P_2}{P_1} - \Bigg(\frac{P_2}{P_1}\Bigg)_\text{res}}{\Bigg(\frac{P_2}{P_1}\Bigg)_\text{res}}, \label{eq:Delta}
\end{equation}
where the quantities with ``res" subscripts denote the integer ratio corresponding to the particular MMR. 
We note that, by convention \citep{Lithwick12}, $\Delta$ tracks the deviation of $n_p/n$ rather than $n/n_p$ (Eq.\:\ref{eq:Delta}), leading to the sign flip on the rightmost axis of Fig.\:\ref{fig:qualitative}. 

Continuing with the dynamical evolution, conjunctions 4-6 occur to the right of the line of apses, and thus slow down the test particle, slowing the advance of the location of successive conjunctions until the test particle's mean motion drops back down below the resonant value after conjunction 6 (right panel).
The next few conjunctions (not pictured in the left panel) still happen to the right of the line of apses, so the test particle's mean motion now continues dropping below the resonant value, and the conjunction location sweeps back, oscillating back and forth, similar to a pendulum (right panel).

Indeed, the dynamics of a pendulum provides a powerful model for resonance \citep[e.g.,][]{Murray99, Tremaine23}, and the analogy becomes exact when the eccentricities and pericenter orientations are approximated as constant.
In this paper we will focus on this so-called ``pendulum approximation", which is generally good as long as the eccentricities are not too small, as quantified in Sec.\:\ref{sec:limits}.

\section{The Geometry of Resonance}

\subsection{The Conjunction Angle} \label{sec:conjunctionangle}

From the qualitative discussion above, the azimuthal location at which a conjunction occurs determines how much it increases or slows down the test particle's mean motion.
It is therefore valuable to consider a conjunction angle anchored to the orbital geometry of the problem (rather than an arbitrary reference direction).
We choose to measure the azimuthal angle where conjunction occurs from the location where the interplanetary separation is minimum (for technical reasons this will later simplify the results). 
In terms of standard orbital angles, for the case above this can be expressed as $\lambda_{\rm{conj}} - \varpi$, where $\lambda_{\rm{conj}}$ is the mean longitude at which conjunction occurs\footnote{Technically, conjunctions occur when the true longitudes are equal and the planets line up. However, at the small eccentricities and leading-order approximation we consider, we can ignore the small additional corrections between the mean longitudes $\lambda$ and the true longitudes $f$. We consider this correction in Appendix \ref{sec:erelHill}.}, and $\varpi$ is the longitude of pericenter of the outer, eccentric orbit.
For example, we saw in Sec.\:\ref{sec:qualitative} that the equilibrium configuration is at $\lambda_{\rm{conj}} - \varpi = \pi$ when conjunctions happen at apocenter, where the two orbits are most widely spaced.

One difficulty with this definition is that discrete conjunctions are challenging to manipulate mathematically.
We can therefore define a continuous ``conjunction angle" $\theta$ as
\begin{equation}
\theta \equiv \frac{p\lambda - (p-q)\lambda_p - q\varpi}{q} \stackrel{at\:conjunction}{=} \lambda_{\rm{conj}} - \varpi. \label{eq:theta}
\end{equation}
In principle the two definitions disagree, but they are equivalent at conjunction when the longitudes are equal, as one can verify by substituting $\lambda=\lambda_p=\lambda_{\rm{conj}}$ above.
One can therefore think of $\theta$ as ``smoothing out" the discrete changes in the conjunction angle visible in the left panel of Fig.\:\ref{fig:qualitative}. 
The angle $\theta$ would initially correspond to the angular distance from pericenter to Conjunction 1, and then move smoothly counterclockwise, lining up with Conjunction 2 at the time of the second conjunction.

\subsection{Phase Space Plots} \label{sec:confspace}

\begin{figure}
    \centering
    \resizebox{0.99\columnwidth}{!}{\includegraphics{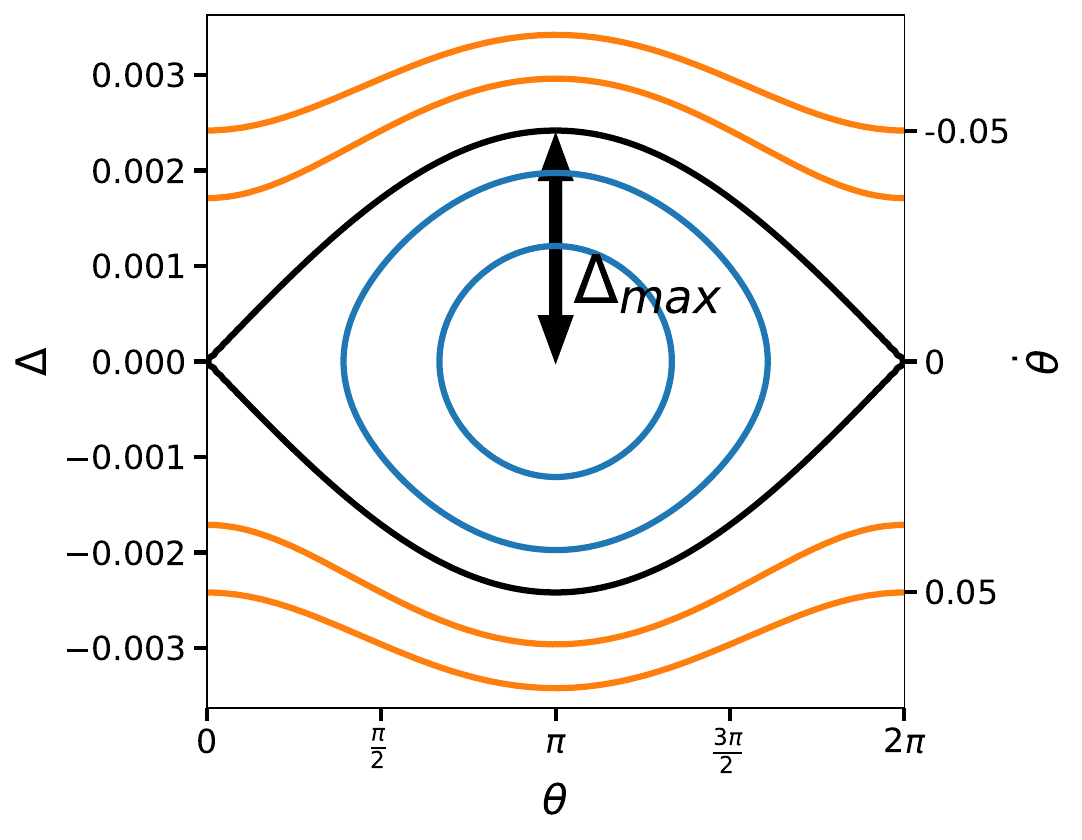}}
    \caption{Smoothed evolution of the fractional deviation from resonance $\Delta$ vs. the angle of conjunction $\theta$, where each curve corresponds to the trajectory followed starting from a different initial condition (using the analytical model summarized in Sec.\:\ref{sec:summary}. Blue curves correspond to initial conditions that are in resonance, where the conjunction angle $\theta$ oscillates within a limited range. Orange curves correspond to configurations out of resonance where the conjunction angle circulates across all values from $[0, 2\pi]$. The black curve is the separatrix trajectory separating resonant librating solutions from non-resonant circulating solutions. The width of the resonance $\Delta_\text{max}$ is the maximum value of $\Delta$ that can remain in resonance, measured where the separatrix is widest. The stellar mass, inner planet mass and the initial eccentricity and pericenter of the outer test particle are always fixed to the values used in Fig.\:\ref{fig:qualitative}, and the innermost blue curve corresponds to the particular trajectory plotted in the right panel of Fig.\:\ref{fig:qualitative}.
    \label{fig:phasespace}}
\end{figure}

If we smooth out the spikes visible in the right panel of Fig.\:\ref{fig:qualitative} to obtain a continuous sinusoid and, instead of plotting the evolution of the period-ratio's deviation from resonance $\Delta$ vs. time, we track the evolution of $\Delta$ vs. the conjunction angle $\theta$, we obtain the innermost blue curve in Fig.\:\ref{fig:phasespace} (approximately a circle).
We see that $\Delta$ oscillates between approximately $\pm 10^{-3}$ (right panel of Fig.\:\ref{fig:qualitative}), while the conjunction angle $\theta$ deviates by about $\pi/3$ from the equilibrium, as seen in the left panel of Fig.\:\ref{fig:qualitative}. 

Because one is often interested in oscillations in the period ratio around the resonant value, it is tempting to think of $\Delta$ as analogous to the pendulum's angle as it oscillates back and forth.
This is imprecise; $\Delta$ corresponds instead to the pendulum's angular velocity (which also oscillates), plotted along the right axis of Fig.\:\ref{fig:phasespace}.
Indeed, if $\Delta = 0$, then the next conjunction happens at the same angular location, so $\dot{\theta}=0$; when the period ratio is not at the resonant value (non-zero $\Delta$), the conjunction angle varies and has a non-zero time derivative.
Like the pendulum's angle and velocity, the conjunction angle $\theta$ and $\Delta$ repeatedly trade off against one another.

Phase space plots like Fig.\:\ref{fig:phasespace} are useful for building up a picture of the different dynamical behaviors for several initial conditions, each of which sweep out their own curve.
Lines in blue correspond to initial conditions that are trapped in the resonance; $\Delta$ oscillates around zero, trading off against the conjunction angle, which ``librates" (oscillates) in a confined range around the stable equilibrium at $\pi$, like a pendulum rocking back and forth.
Lines in orange correspond to configurations outside the resonance; these start at a value of $\Delta$ that is too large to return to zero (given the adopted masses and eccentricities) and where the conjunction angle circulates over the full range between $[0, 2\pi]$.
This is analogous to a pendulum with ``too much" initial kinetic energy that repeatedly goes over the top.
The special trajectory separating librating from circulating solutions is called the separatrix and is plotted in black.

We are often interested in calculating the ``width" of the resonance i.e., the maximum value $\Delta_\text{max}$ that can still remain in resonance; this corresponds to the value on the separatrix at the location where the resonance is widest (see Fig.\:\ref{fig:phasespace}).
This is useful not only for quantifying the size of the resonant region (e.g., does the 3:2 MMR matter for a pair of exoplanets at $\Delta = 0.01$ given their masses and eccentricities?), but also for checking whether the widths of adjacent MMRs overlap one another, which is a useful heuristic criterion for chaotic behavior \citep{Chirikov79}.

The size of the resonant region ultimately depends on the strength of the gravitational interaction between the pair of planets relative to that with the central star. 
It should therefore depend both on the planet-star mass ratio, as well as on the (normalized) eccentricity, which determines the asymmetry pre and post conjunction (Sec.\:\ref{sec:qualitative}). We explore these relationships in the following sections.

\subsection{Higher Order MMRs} \label{sec:higherintro}

\begin{figure}
    \centering
    \resizebox{0.99\columnwidth}{!}{\includegraphics{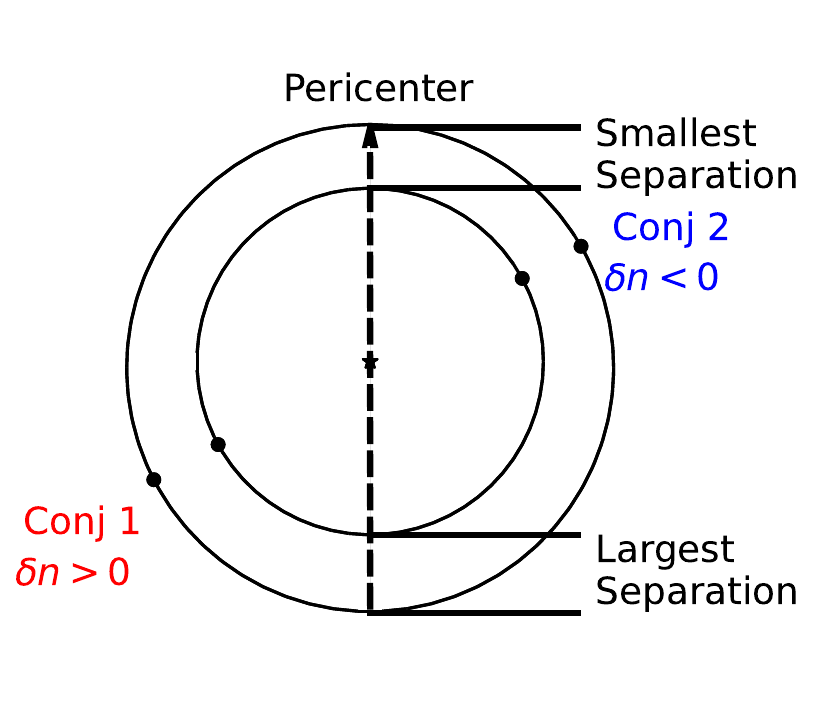}}
    \caption{Analogous geometry to the left panel of Fig.\:\ref{fig:qualitative}, but for a second-order (5:3) MMR that undergoes two conjunctions per cycle. The fact that the planets orbit at nearly constant rates implies that if one conjunction happens at a particular location (e.g., Conj 1 in red), then the other conjunction in the cycle must happen on the opposite side (Conj 2, blue). In this paper we explore the idea that this leads to a cancellation of their effects at leading order, explaining why higher-order MMRs are weaker. See text for details.
    \label{fig:second}}
\end{figure}

While first-order MMRs of the form $p$:$p-1$ have the strongest effects, we also want to explore why higher order $p$:$p-q$ MMRs with $q>1$ are weaker.
For a first-order MMR, the inner planet executes one more orbit than the inner planet per cycle, resulting in a single conjunction.
For, e.g., the second-order 5:3 MMR, the inner planet would make five orbits in the time the outer body does three, which means the inner planet had to overtake twice. 
The order of the resonance $q$ therefore corresponds to the number of conjunctions before the cycle repeats.
%For a $p:p-q$ MMR, the inner planet orbits $p$ times in the time it takes the outer planet to orbit $p-q$ times, and then the cycle repeats.
%This means that the inner planet overtakes $q$ times each cycle.%, so each cycle lasts $q t_{conj}$.

In the left panel of Fig.\:\ref{fig:second}, we show a plot analogous to the left panel of Fig.\:\ref{fig:qualitative}, now considering a second-order 5:3 MMR. 
Because the planetary perturbations are small and the planets orbit at approximately constant rates, the two conjunctions in the cycle must be approximately equally spaced in time.
If, starting at conjunction, the next cycle repeats after five and three inner and outer orbits, respectively, then the second conjunction must occur halfway through the cycle (after 5/2 and 3/2 inner and outer orbits, respectively) when the planets are on the opposite side of their orbits (Fig.\:\ref{fig:second}).
In general for a $q$th order MMR, we can then think of $\theta$ (Eq.\:\ref{eq:theta}) as measuring where the first conjunction happens in the cycle, with the remaining conjunctions occurring at equally spaced intervals, i.e., conjunctions at $\theta + 2\pi k/q$, for $k=0,...,q-1$.
Thus, for example in the top left panel of Fig.\:\ref{fig:folding}, a given initial condition would follow a single curve, e.g., one of the approximately circular paths around $\theta = \pi/2$.
This would track the variations in the location of \textit{one} of the two conjunctions in the cycle, and one could obtain the second conjunction location by adding $\pi$.

The idea we pursue in this paper is that the physical reason why a second-order MMR is weaker than a first-order MMR is that the effects from the two conjunctions are almost equal and opposite, leading to a partial cancellation.
We explore this idea in Sec.\:\ref{sec:higher}.

\subsection{The Resonant Angle} \label{sec:resangle}

\begin{figure}
    \centering
    \resizebox{0.99\columnwidth}{!}{\includegraphics{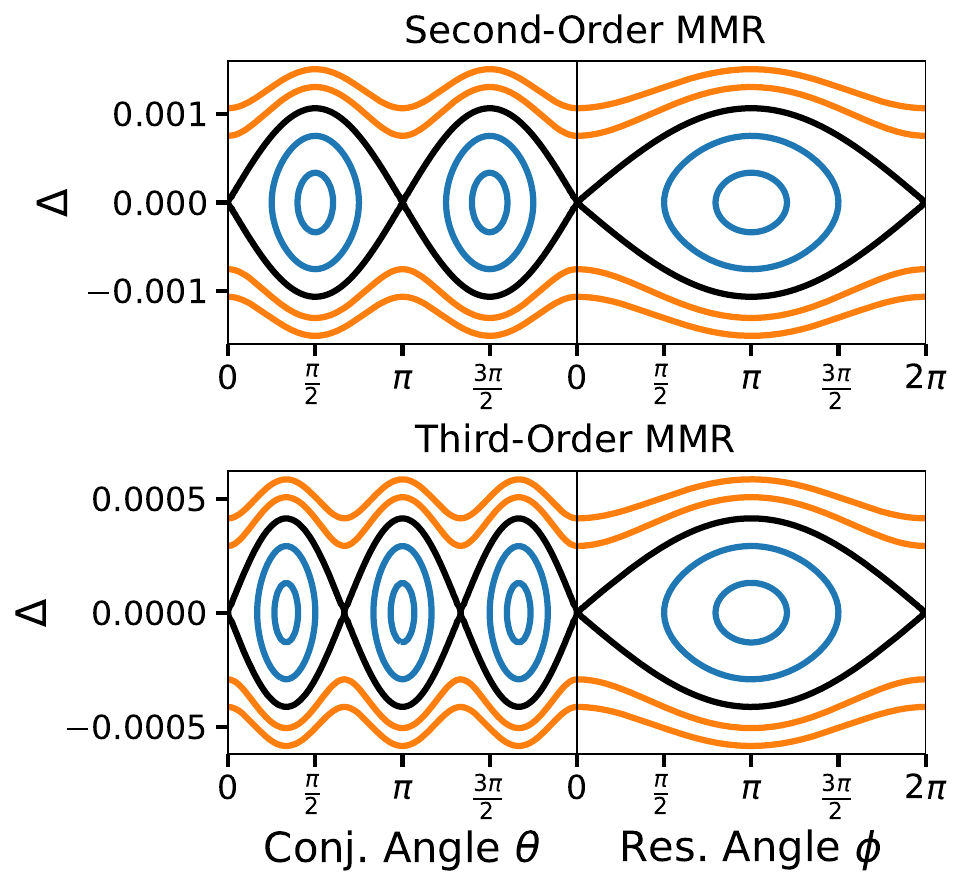}}
    \caption{Analogous plots to Fig.\:\ref{fig:phasespace}, but for second and third-order MMRs, highlighting their $q$-fold symmetry (where $q$ is the order of the resonance). The left panels plot $\Delta$ vs. $\theta$, where $\theta$ can be thought of as measuring where the first conjunction occurs in the cycle of $q$ conjunctions (Fig.\:\ref{fig:second}). The right panel instead plots $\Delta$ vs. the resonant angle $\phi \equiv q\theta$, which acts to fold the $q$ islands into a single one with the shape of a first-order MMR, with the dynamics of a simple pendulum.
    \label{fig:folding}}
\end{figure}

While the conjunction angle $\theta$ provides a clear physical intuition, it is mathematically useful for higher order MMRs to introduce the resonant angle \citep[e.g.,][]{Murray99},
\begin{equation}
\phi \equiv q\theta. \label{eq:phi}
\end{equation}
This acts to ``fold" the $q$ conjunctions per cycle onto one another, which in the pendulum approximation leads to a common dynamical model for MMRs of all orders, as seen in Fig.\:\ref{fig:folding}.

As discussed in detail in Sec.\:\ref{sec:higher}, this transformation causes the two resonant islands for a second-order MMR (top left panel) and the three resonant islands for a third-order MMR (bottom left panel), to get mapped into a single island (corresponding right panels).
In the pendulum approximation considered in this paper (see Sec.\:\ref{sec:limits} for the limits of this approximation), the resulting island has the same shape as a first-order MMR (Fig.\:\ref{fig:phasespace}), and trajectories follow the dynamics of a simple pendulum.

We can also see that the stable equilibrium configuration always corresponds to a period ratio at the resonant value ($\Delta = 0$, Eq.\:\ref{eq:Delta}), and resonant angle\footnote{We note that this convenient property is due to the fact that we defined $\theta$ to be measured from the location where the orbits are \textit{closest} to one another. Measuring it from the location where they are farthest apart \citep[e.g.,][]{Murray99} means that the equilibrium resonant angle is instead 0 for odd-order MMRs, and $\pi$ for even-order MMRs.} 
\begin{equation}
\phi_{eq} = \pi. \label{eq:phieq}
\end{equation}

This can be combined with the fact that the $q$ conjunctions in a $q$th order MMR have to be equally spaced to work out that, e.g., for a third-order MMR, the first conjunction in the stable equilibrium configuration must lie at $\theta_\text{eq}=\phi_\text{eq}/q=\pi/3$, with the remaining ones $2\pi/q$ apart.
This matches the bottom left panel of Fig.\:\ref{fig:folding}, with equilibria at $\pi/3$, $\pi$ and $5\pi/3$. 

\section{Summary} \label{sec:summary}

We begin by summarizing the main results, both as a roadmap for the paper, and for ease of reference.
Many of these results should not be obvious on a first reading and we will explore them in more detail in subsequent sections.
For clarity in these general results for two massive planets, in this section we will refer to the inner planet with subscript one, and its outer neighbor with subscript 2.

In the traditional celestial mechanics approach to MMRs, one isolates the appropriate resonant harmonics in a Fourier expansion of the gravitational potential between a pair of planets (Appendix \ref{sec:disturbing}).
While this is a powerful mathematical approach, it leads to several technical inconveniences.
First, the resonant dynamics depend on complicated combinations of the planet-star mass ratios $\mu_i$ and eccentricity vectors ${\bf e_i}$ (with magnitude given by the $i$th planet's orbital eccentricity, and pointing toward pericenter).
Second, the corresponding amplitudes are a function of the semimajor axes; they are therefore different for each particular MMR, and they vary by orders of magnitude for MMRs at close orbital separations.

\cite{Hadden19} realized that the well known dynamical properties of the compact (Hill) limit \citep{Henon86, Namouni96} implied that both these issues should disappear.
First, he shows \citep[see also][]{Hadden18} that in this regime there is a powerful mapping from the general problem of two massive planets near MMRs (with several subresonances) to the much simpler case of the PCR3BP (with a single resonance), summarized below.
An equivalent result (though not stated in this way) was known independently for first-order MMRs \citep{Sessin84, Wisdom86, Deck13}, but \cite{Hadden19} presents a general result for arbitrary-order MMRs at close separations through canonical perturbation theory.
In this paper we instead explore them through complementary physical arguments in Sec.\:\ref{sec:general}; we summarize the mapping for reference in Table \ref{tab:mapping}.

This result significantly reduces the number of parameters and dynamical variables to consider.
For example, in the general problem, there is only one linear combination of the eccentricity vectors that matters for the resonant dynamics, called the ``relative eccentricity" \citep[e.g.,][]{Namouni96}
\begin{equation}
{\bf e_{12}} = {\bf e_2 - e_1}. \label{eq:erel}
\end{equation}

\begin{figure*}
    \centering
    \resizebox{0.99\textwidth}{!}{\includegraphics{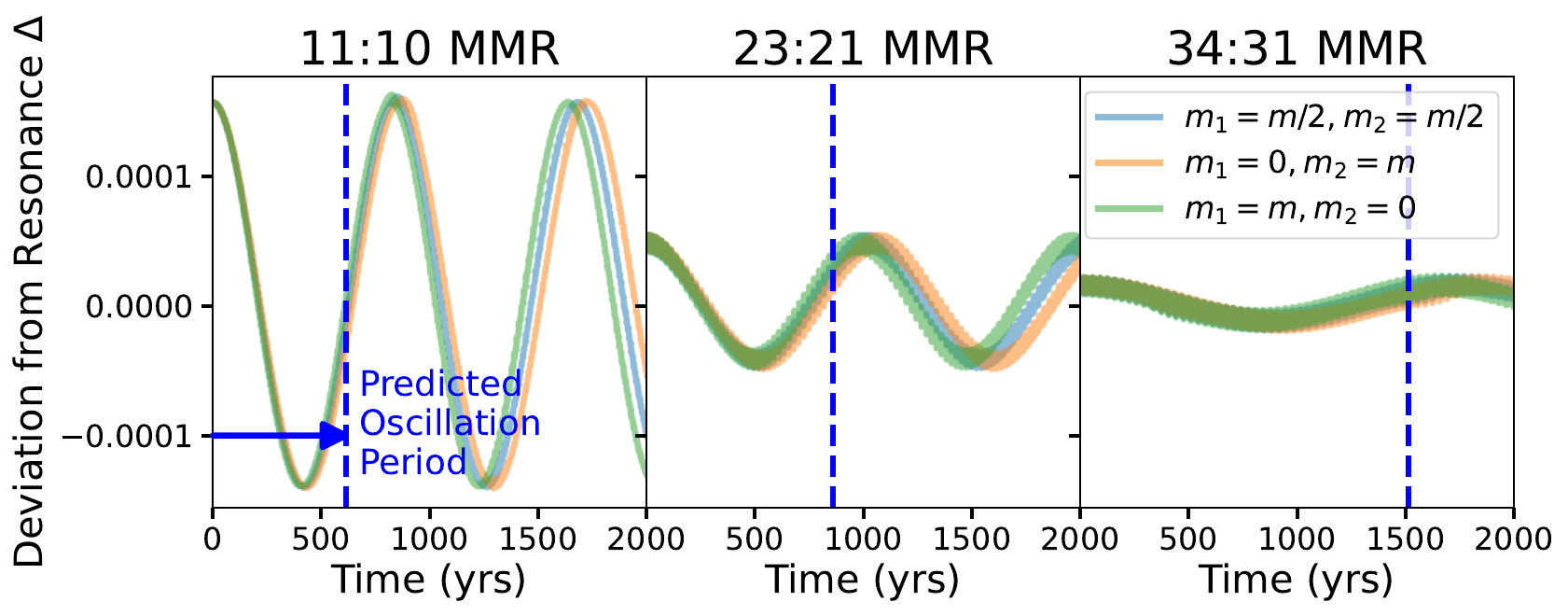}}
    \caption{Period-ratio deviation from resonance $\Delta$ vs. time for a first (left panel), second (middle panel) and third (right panel) MMR. The strong agreement between the three colors in each panel demonstrates the approximate mapping between the general (blue) and test-particle (green and orange) cases detailed in Table \:\ref{tab:mapping} (see text for details). The dashed blue lines show the oscillation period corresponding to the frequency given in Eq.\:\ref{eq:frequency}. In order to more cleanly see the oscillations for the second and especially the third-order MMR, we use the \texttt{celmech} package \citep{celmech} to calculate mean orbital periods as a function of time.
    \label{fig:frequencies}}
\end{figure*}

We provide an illustration in Fig.\:\ref{fig:frequencies}, showing approximately the same dynamical evolution for three different setups (colored curves). 
All configurations are co-planar, and planets are initialized such that their first conjunction occurs at the location where the orbits are most widely spaced ($\theta=\pi$).

The green curves are initialized with all the mass ($\mu=10^{-7}$) in the inner planet on a circular orbit, and all the eccentricity $\approx 0.01$ in the outer, massless body's orbit.
The orange curves instead put all the mass in the outer body, and all the eccentricity in the inner body, while the blue curves split the mass and eccentricity evenly, with pericenters antialigned (so the relative eccentricity $e_{12} \approx 0.01$ in all three cases).
In the three panels we select a first, second and third order MMR at close orbital separations and similar period ratios (1.1, 1.095, 1.097, respectively).
In all cases, the three colored curves nearly agree, illustrating that one can approximately map the general co-planar MMR dynamics to either of the much simpler test-particle cases through Table \ref{tab:mapping}.

The correspondence worsens somewhat at wider separations, but the approximation is still useful for many applications, and can provide intuition and a vocabulary for thinking about more complicated combinations of variables valid at wider separations that are derived through canonical perturbation theory \citep[e.g.,][]{Deck13, Hadden19}.

\begin{table*}
    \centering
    \begin{tabular}{llll}
        Name & General Case & Inner Massive Circ. Orb. & Outer Massive Circ. Orb.\\
         &  & Outer Test Particle & Inner Test Particle\\
        Period Ratio Deviation & $\Delta$ (Eq.\:\ref{eq:Delta}) & $-\frac{n_2 - n_{2,\text{res}}}{n_{2,\text{res}}}$ & $\frac{n_1 - n_{1,\text{res}}}{n_{1,\text{res}}}$\\
        Mass Ratio & $\mu = \frac{m_1+m_2}{M_\star}$ & $\mu = \frac{m_1}{M_\star}$ & $\mu = \frac{m_2}{M_\star}$ \\
        Eccentricity & $e_{12} = |{\bf e_2} - {\bf e_1}|$ & $e_2$ & $e_1$\\
        Min. Sep. between Orbits & $\varpi_{12}$ = arg(${\bf e_{12}}$) & $\varpi_2$ & $\varpi_1 + \pi$ \\
        Conjunction Angle $\theta$ & $p\lambda_2 - (p-q)\lambda_1 - q\varpi_{12}$ & $p\lambda_2 - (p-q)\lambda_1 - q\varpi_{2}$ & $p\lambda_2 - (p-q)\lambda_1 - q(\varpi_{1} + \pi)$ \\
        Constant ``clock" & $n_\text{COM}$ & $n_1$ & $n_2$\\
         &  &  & \\
    \end{tabular}
    \caption{Mapping of variables in the compact limit between the general case of two massive planets on co-planar, eccentric orbits, and the co-planar, circular restricted three-body problem (in the cases where the inner or outer body is massless). See Appendix \ref{sec:spacings} for leading-order corrections at wider spacings.}
    \label{tab:mapping}
\end{table*}

Second, a given relative eccentricity $e_{12}$ might be negligible at wide separations, but be big enough for orbits to cross for closely spaced orbits.
\cite{Hadden18} therefore define an eccentricity $\tilde{e}$ normalized to the value at which the orbits would cross
\begin{equation}
\tilde{e} \equiv \frac{e_{12}}{e_c}. \label{eq:etilde}
\end{equation}
The crossing eccentricity $e_c$ corresponds to when the radial excursions of size $\approx ae$ become equal to the orbital separation $\Delta a = a_2 - a_1$.
For planets near a $p:p-q$ MMR, Kepler's third law then implies that for closely spaced orbits with small $\Delta a/a$, 
\begin{equation}
e_c \equiv \frac{\Delta a}{a} \approx \frac{2}{3} \frac{\Delta P}{P} \approx \frac{2}{3}\Bigg(1-\frac{p-q}{p}\Bigg) = \frac{2q}{3p}, \label{eq:ecross}
\end{equation}
where we have approximated the test particle's orbital period as being at the resonant value.

\cite{Hadden18} show that by expressing things in terms of normalized ($\tilde{e}$) rather than absolute ($e$) eccentricities, the relative strengths of MMRs at different separations are always of order unity (instead of becoming very large for MMRs corresponding to orbits with close separations, see \citealt{Murray99}).
While in the traditional approach there is a different strength for each $p$:$p-q$ MMR \citep{Murray99}, this normalization leads to coefficients that are universal to all MMRs of the same order $q$ \citep{Hadden18}.
We explore this simplification in Sec.\:\ref{sec:normalizedMMRs}, and summarize its implications for MMR widths and oscillation frequencies in the next subsection.

%a classical result is that in this Hill limit, the dynamics only depends on the relative motion \citep[e.g.,][]{Henon86}, and in particular on the relative eccentricity vector \citep{Namouni96}
%\begin{equation}
%{\bf e_{12}} = {\bf e_2} - {\bf e_1},
%\end{equation}
%where the eccentricity vector ${\bf e_i}$ has a magnitude given by the $i$th planet's orbital eccentricity, with direction specified by its corresponding pericenter.
%\cite{Hadden19} shows that in the compact limit, different subresonances involving varying combinations of the eccentricities all collapse into a single resonance involving only ${\bf e_{12}}$ (Sec.\:\ref{sec:collapse}).
%We show in Sec.\:\ref{sec:erel} that this powerful result has a simple geometric interpretation.

%Similarly, the dynamical evolution does not depend on the two orbits' mean motions independently, but only on their difference, or equivalently $\Delta$
%Additionally, in the Hill limit, the dynamical evolution does not depend on the two planetary masses individually, but only on their sum (Sec.\:\ref{sec:mass}.

%Furthermore, the fact that only a single linear combination of the eccentricity vectors enters the dynamics implies that one can map the general problem of 2 massive pl

\subsection{MMR Widths and Frequencies} \label{sec:widths}

The results from \cite{Hadden18} and \cite{Hadden19} imply (see Sec.\:\ref{sec:widthsgen}) that in the pendulum approximation, the resonant angle $\phi$ near a $q$th-order MMR follows the dynamics of a simple pendulum with Hamiltonian

\begin{equation}
H = \frac{\dot{\phi}^2}{2} + \omega^2  \cos \phi. \label{eq:Hgen}
\end{equation}
where the frequency for small-amplitude oscillations is given by
\begin{equation}
\omega = q A_q \frac{\sqrt{\mu \tilde{e}^q}}{e_c} n. \label{eq:frequency}
\end{equation}
The $A_q$ are coefficients of order unity that are universal to all MMRs of a given order $q$, and in the closely spaced limit, $n$ could be taken as either planet's mean motion\footnote{It is mathematically cleanest to identify $n = n_\text{COM}$ (see Sec.\:\ref{sec:ncom}), which means that the three different colored curves in each panel of Fig.\:\ref{fig:frequencies} would have a (slightly!) different predicted period, but these corrections are not meaningful at our level of approximation. In practice, given that unmodeled higher harmonics not captured by the pendulum approximation always lead to slower oscillation periods (see text), one will always be closer to the true oscillation period by taking the slower (i.e., the outer) of the two mean motions as $n$.}.
Numerical values for the $A_q$ are listed in Table \ref{tab:Aq}.

The corresponding oscillation periods are plotted as dashed blue lines in Fig.\:\ref{fig:frequencies} and closely match the integrations, particularly for the higher order MMRs.
The discrepancy ($\approx 20\%$ too short for the first-order MMR) is mostly due to the effects of MMRs corresponding to higher order multiples of the same integer ratio, e.g., the 6:4 and 9:6 MMRs for the 3:2 MMR, and cannot be incorporated into a simple pendulum model \citep[see the discussion below Fig. 3 in ][]{Hadden19}. 

The integrations in Fig.\:\ref{fig:frequencies} were all initialized with conjunction occurring at the equilibrium location where the orbits are farthest apart, but with a period ratio offset from the resonant value ($\Delta > 0$). 

For this starting choice of $\phi=\pi$, the planet pair would be in resonance for initial values of $\Delta$ out to the width of the MMR (see Fig.\:\ref{fig:phasespace}), which we show below is given by
\begin{equation}
\Delta_\text{max} = 3 A_q \sqrt{\mu \tilde{e}^q}, \label{eq:width}
\end{equation}
for any $q$th-order MMR in the limit of close orbital separations.

\begin{table}
    \centering
    \begin{tabular}{c|c}
        MMR Order & $A_q$\\
        1 & 0.845 \\
        2 & 0.754 \\
        3 & 0.748 \\
        4 & 0.778 \\
        5 & 0.832 \\
        6 & 0.904 \\
        7 & 0.995 \\
        8 & 1.104 \\
        9 & 1.235 \\
        10 & 1.388
    \end{tabular}
    \caption{Numerical prefactors $A_q$ for MMRs of order $q$. They come from the leading-order term in a Taylor expansion of MMR strengths in powers of the normalized eccentricity $\tilde{e}$ \cite{Hadden18}. For details, see Sec.\:\ref{sec:collapse} (see also Sec.\:\ref{sec:powerseries} for how they are related to an expansion of the kick to the mean motion at conjunction in powers of $\tilde{e}$ \citealt{Namouni96}). The particular form we choose for the $A_q$ in this paper was chosen so as to simplify the expressions for the MMR widths and frequencies one is often most interested in.}
    \label{tab:Aq}
\end{table}

\begin{figure*}
    \centering
    \resizebox{0.99\textwidth}{!}{\includegraphics{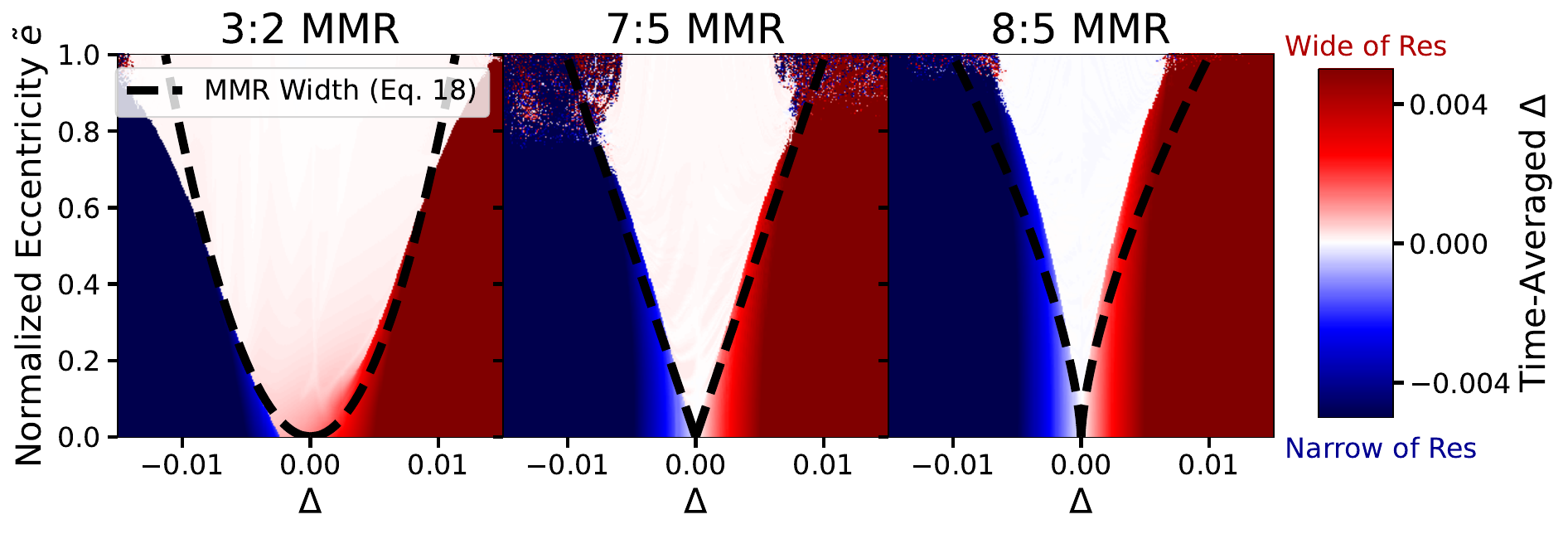}}
    \caption{Grid of N-body simulations of two $m=10^{-5}$ planets around a 1 $M_\odot$ star, for a range of period-ratio deviations from resonance $\Delta$ (Eq.\:\ref{eq:Delta}) and normalized eccentricities $\tilde{e}$ (Eq.\:\ref{eq:etilde}). See text for details. Color shows the time-averaged value of $\Delta$ over 400 outer-planet orbits. Configurations wide of resonance have time-averaged values of $\Delta > 0$ (red), while ones narrow of resonance have mean $\Delta < 0$ (blue). Configurations in resonance oscillate around the resonant period ratio and have time-averaged $\Delta \approx 0$ (white). Dashed black curves show the MMR widths given by Eq.\:\ref{eq:width}.
    \label{fig:widths}}
\end{figure*}

With an important correction for the specific case of the 2:1 MMR (see Appendix \ref{sec:21}), this expression works well for MMRs at close separations (see Appendix \ref{sec:spacings} for estimates of the small corrections out to spacings as wide as the 3:2 MMR), as illustrated in Fig.\:\ref{fig:widths}.
For the MMR in each panel, we initialize a grid of orbital configurations with two $m=10^{-5} M_\odot$ planets around a $1 M_\odot$ star, with a range of normalized eccentricities along the y-axis.
The inner planet always has an orbital period of 1 year, and the outer planet is initialized with periods spanning a range of deviations $\Delta$ from the corresponding MMR along the x-axis.
We then run each configuration for 400 orbits of the outer planet, and plot the time-averaged value of $\Delta$ as a color. 

Configurations that are in resonance will oscillate around the resonant value $\Delta = 0$ and appear white, while pairs that are wide of resonance will remain at a time-averaged value of $\Delta > 0$ (red), and ones narrow of resonance will remain at $\Delta < 0$ (blue).
The width predicted by Eq.\:\ref{eq:width} is shown as a dashed black curve in each panel.

Finally, we note that each integration in Fig.\:\ref{fig:frequencies} was initialized half-way to the separatrix, with $\Delta = \Delta_{max}/2$.
We therefore see from Eq.\:\ref{eq:width} that while the widths of the MMRs decrease significantly, by a factor of $\tilde{e} = 0.15$ between the left ($\propto \tilde{e}^{1/2}$) and right ($\propto \tilde{e}^{3/2}$) panels, the oscillation frequencies are more similar to one another, partially offset by the factor of $q$ in Eq.\:\ref{eq:frequency}.

\section{Scalings for First-Order MMRs in the Hill Limit} \label{sec:scaling}
In this section we provide physical scaling arguments for the widths and libration frequencies of first-order MMRs in a ``smoothed" model for the right panel of Fig.\:\ref{fig:qualitative}, where we ignore the sharp spikes and only consider the net exchange of energy at each conjunction.
Since we have seen in Sec.\:\ref{sec:summary} that the general case of two massive planets on eccentric orbits can be mapped to the circular restricted case, we return to the simple case of Sec.\:\ref{sec:qualitative} where the massive inner planet (with subscript $p$) is on a circular orbit, and the test particle (with no subscripts) is on an eccentric outer orbit. 

Our strategy will be to estimate how these changes to the test particle's mean motion scale with the orbital parameters and conjunction location (Sec.\:\ref{sec:kick}), and how the mean motion in turn determines the location of the next conjunction (Sec.\:\ref{sec:dynamics}).
In Sec.\:\ref{sec:pendulum} we will then couple these results to obtain the dynamics of a simple pendulum.
We begin with some useful preliminaries in the Hill limit where the orbits are closely spaced.

\subsection{The Hill Limit} \label{sec:Hill}

\begin{figure}
    \centering \resizebox{\columnwidth}{!}{\includegraphics{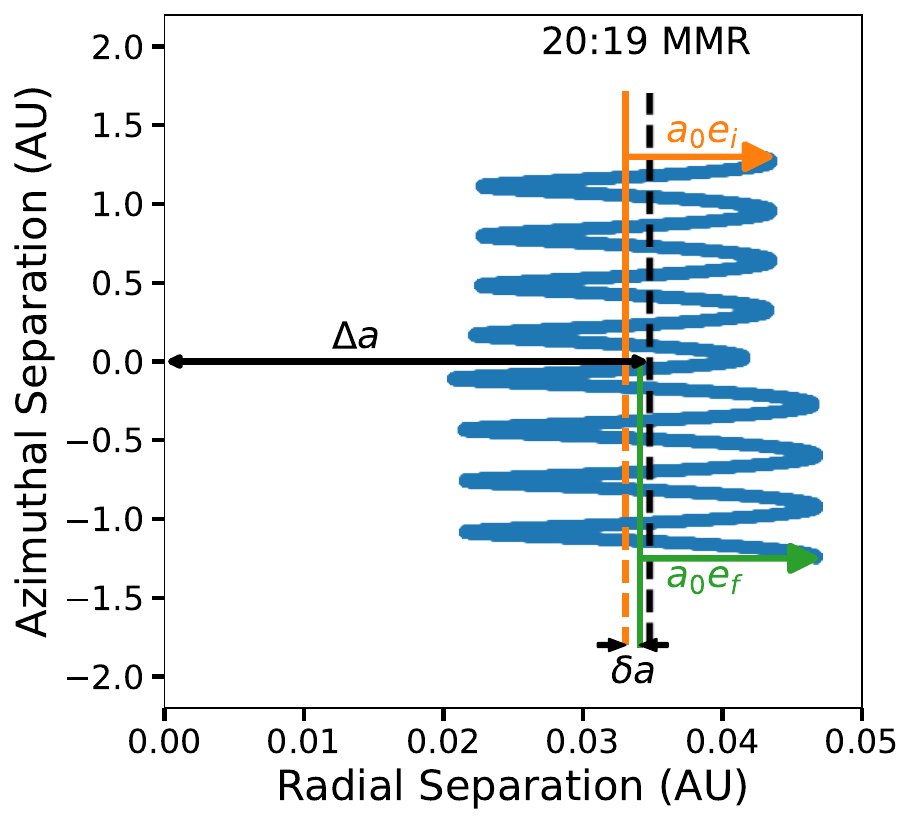}}
    \caption{View of a close encounter in Hill's coordinates, which rotate at a uniform rate with the inner planet, so it remains fixed at (0,0). The test particle starts ahead of the planet along its orbit, at positive y. Because of the close orbital separation near the 20:19 MMR (black dashed line), it takes many orbits for the inner planet to overtake the test particle. As it moves between pericenter and apocenter, the test particle therefore undergoes several radial oscillations around the initial semimajor axis (vertical orange line) before the close encounter. When it reaches conjunction at y=0, the semimajor axis changes nearly instantaneously by $\delta a$ to the solid green line, and the test particle continues on an effectively unperturbed orbit until the next conjunction.
    \label{fig:encounter}}
\end{figure}

In the compact or Hill limit, there are two small quantities,

\begin{equation}
\frac{\Delta a}{a} \ll 1 \:\:\:\:\:\: e \ll 1,
\end{equation}
where compact implies a small fractional separation between the orbits $\Delta a \equiv a - a_p$, and the eccentricities must also be small in order for the orbits not to cross.

It is useful to consider the problem in a frame rotating uniformly with the inner planet, which we assume orbits on an unperturbed, circular orbit at a rate $n_p$.
In rotating polar coordinates, the inner body then remains fixed at $(r=a_p, \theta'=0)$ (we use $\theta'$ to distinguish from the conjunction angle $\theta$).
To track the test particle in this rotating frame, we define a radial separation $x=r-r_p$ and an azimuthal separation $y=a_p\theta'_{rel}$, where $\theta'_{rel}$ is the angle between their position vectors.
In the limit where $\Delta a \rightarrow 0$, the curvature in these coordinates becomes negligible, so the Hill problem adopts Cartesian coordinates to simplify the equations of motion \citep[e.g.,][]{Tremaine23}.

In Hill's coordinates, the inner massive planet therefore remains at $(x_p,y_p) = (0,0)$, while a slower-moving, unperturbed exterior body on a circular orbit would drift downward on a vertical line with fixed $x$ (e.g., along the dashed black line with fixed $x = \Delta a$ in Fig.\:\ref{fig:encounter}).
By contrast, an unperturbed eccentric orbit (blue) will move closer and farther from the inner planet by $\approx ae$ as it moves between pericenter and apocenter every orbit.

Figure \ref{fig:encounter} shows the case of an eccentric test particle near the 20:19 MMR corresponding to a semimajor axis of $a_0$ (dashed black line).
The test particle starts at the top of the figure (blue) with an initial orbital eccentricity $e_i$, and executes several effectively unperturbed radial excursions of size $a_0 e_i$ along a Keplerian orbit (because the eccentricities are already small, we can ignore the small deviations of the actual semimajor axis from the resonant value $a_0$).
Conjunction occurs at an azimuthal separation of zero.
For closely spaced orbits, the interplanetary separation at conjunction is so much smaller than it is between conjunctions (see left panel of Fig.\:\ref{fig:qualitative}), that the test particle receives an approximately instantaneous kick onto a new Keplerian orbit with elements $a_f$ and $e_f$ that it will then follow until the next conjunction.

We note that because closely spaced orbits have similar mean motions, there can be many orbits between successive conjunctions (e.g., for the 10:9, the inner planet does one more orbit than the outer one before overtaking again, so there will be 10 inner-planet orbits between conjunctions).
This introduces two important timescales for the planetary encounter.

%\footnote{Because the eccentricities are always assumed small, we can treat $a$ as constant when working to leading order, since the small corrections to $a$ would only lead to terms at quadratic order in small parameters.}.
%Because we assume that the test particle's semimajor axis is close to a $j$:$j-k$ MMR, we expand $x$ as $\Delta a_{res} + \delta a$, where $\Delta a_{res}$ is the resonant semimajor axis (black dashed line in Fig.\:\ref{fig:encounter}, which remains fixed), and $\delta a$ is the deviation from resonance, which varies (slightly) at each successive encounter (see orange vs. green dashed curves in Fig.\:\ref{fig:encounter}).

First, the planets drift apart at a rate $\Delta n = n_p-n$, so the time between conjunctions is given by $t_{conj} = 2\pi / \Delta n$, which increases for more tightly spaced orbits with similar mean motions.
Kepler's 3rd law yields $\Delta n/n \approx \frac{3}{2} \frac{\Delta a}{a} = \frac{3}{2}e_c$, so 
\begin{equation}
t_{conj} = \frac{4\pi}{3 e_c\:n} = \frac{2}{3}\frac{P}{e_c}. \label{eq:tconj}
\end{equation}
A small $e_c$ (at close separations) thus leads to many pericenter and apocenter passages in the time between conjunctions (Fig.\:\ref{fig:encounter}).

Finally, even though the time between conjunctions can be long, the timescale of the encounter is always short.
For small eccentricities, the interplanetary separation reaches a minimum of $\approx \Delta a$ at $y=0$.
The corresponding gravitational force therefore remains roughly constant (within a factor of two) inside an azimuthal separation $|y| \lesssim \Delta a$; beyond that the gravitational force falls off rapidly. 
The characteristic timescale of the interaction $\Delta t_{int}$ is therefore given by the time it takes for $y$ to change by $\Delta a$ in the rotating frame, moving at a rate $\dot{y} \approx a \Delta n$:
\begin{equation}
\Delta t_{int} \sim \frac{\Delta a}{a \Delta n} = \frac{P}{3\pi}, \label{eq:dtint}
\end{equation}
which is independent of orbital separation (and thus approximately true even for modestly spaced MMRs out to the 2:1).
This makes it possible to model the interaction as a sequence of instantaneous kicks to the test particle's mean motion, with the bodies drifting along unperturbed orbits between conjunctions \citep[e.g.,][]{Duncan89}.

Our goal is to predict the kicks (changes) to the semimajor axis $\delta a$ (Fig.\:\ref{fig:encounter}), or equivalently to the mean motion $\delta n$.
To do this simply, we approximate the eccentricity as constant throughout the evolution, neglecting its small fractional variations of typically a few percent. 
It is this choice that renders the resulting evolution identical to that of a pendulum and it is therefore typically referred to as the pendulum approximation \citep[e.g.,][]{Murray99}.
An important situation in which it would break down is when the eccentricity is close to zero, where a kick to larger eccentricity would represent a large fractional change. 
In this case, additionally modeling the eccentricity evolution is important, as discussed in Sec.\:\ref{sec:limits}.

\subsection{A dynamical model} \label{sec:dynamics}

For a test particle near a $p$:$p-q$ MMR, we define a reference mean motion $n_\text{res}$ such that $n_\text{res}/n_p \equiv (p-q)/p$.
In our setup, the inner planet is unperturbed so the inner planet's mean motion is always $n_p$, but $n$ can be offset from $n_\text{res}$.
During the time between conjunctions when the planets do not interact significantly, the mean longitudes move according to their unperturbed mean motions, so if we take a time derivative of Eq.\:\ref{eq:theta} and approximate $\dot{\varpi} = 0$\footnote{This is a good approximation in the pendulum limit we consider in this paper. It becomes problematic at low eccentricities where the pericenter precesses rapidly, and a more complicated model is required (see Sec.\:\ref{sec:limits}).}, $\theta$ varies at a rate \begin{eqnarray}
\dot{\theta} &=& \frac{pn - (p-q)n_p}{q} = \frac{p\left[n_\text{res} + (n -n_\text{res})\right] - (p-q)n_p}{q} \nonumber \\ 
&=& \frac{2}{3}\frac{n - n_\text{res}}{e_c} \approx -\frac{2}{3}\frac{n_\text{res} \Delta}{e_c} , \label{eq:thetadot}
\end{eqnarray}
where we used the resonance condition $pn_\text{res} - (p-q)n_p = 0$, used Eq.\:\ref{eq:ecross} to write $p/q$ in terms of the crossing eccentricity, and in the last approximation expanded $n$ around $n_\text{res}$ in Eq.\:\ref{eq:Delta} to leading order to obtain $\Delta \approx -(n-n_\text{res})/n_\text{res}$. 
The conjunction angle thus remains fixed in time when $n = n_\text{res}$ ($\Delta = 0$), and changes uniformly with time at a rate that becomes larger the further the test particle's mean motion is from resonance\footnote{The Hill problem is well posed independent of whether or not one is close to resonance, and one can build a general mapping from one conjunction to the next \citep{Duncan89, Namouni96}.
But near an MMR, $\Delta$ is small and therefore $\theta$ only changes slightly between conjunctions.
This enables the exploration in this paper of a smoothed model in which the orbits change continuously in time.}.

We have seen that $n$ changes at conjunction, and that these changes are a function of the angle $\theta$ where the conjunctions happen (left panel of Fig.\:\ref{fig:qualitative}), leading to a coupled set of differential equations for $n$ and $\theta$.
We can decouple them by taking a time derivative of Eq.\:\ref{eq:thetadot},
\begin{equation}
\ddot{\theta} = \frac{2}{3} \frac{\dot{n}}{e_c}  \label{eq:thetaddot}
\end{equation}
and finding a way to express $\dot{n}$ as a function of $\theta$. 
%Additionally, we note that since $\delta n = n_2 - n_{20}$, $\delta \dot{n} = \dot{n}_2$.
%To help generalize later to the general case of two massive planets, we point out now that $\delta n$ (the deviation from the resonant mean motion $n_{20}$) differs from $\Delta n$ (the difference between the two mean motions) only by a constant $\Delta n_{res} \equiv n_{10}-n_{20}$
%\begin{equation}
%\delta n \equiv \Delta n - \Delta n_{res},
%\end{equation}
%where in this test-particle setup, the inner planet's mean motion never changes so $n_1 = n_{10}$ always.
%In particular, this implies that $\delta \dot{n} = \Delta \dot{n}$.

In Sec.\:\ref{sec:qualitative}, we argued that the mean motion only changes significantly at conjunction.
However, since the effects only build up over many conjunctions, we can imagine a smoothed out version of the right panel of Fig.\:\ref{fig:qualitative}, omitting the spikes and smearing out the changes in mean motion to act continuously between conjunctions like we did above for $\theta$ (Eq.\:\ref{eq:theta}). 
In this picture, the smoothed time derivative of $n$ is then given by the size of each kick $\delta n$ divided by the time between conjunctions,
\begin{equation}
\dot{n} \approx \frac{\delta n}{t_{conj}} \approx \frac{3}{4\pi}e_c n \delta n %\frac{\mu \tilde{e}}{e_c} n^2 \sin \phi,        
\end{equation}
where we have used Eq. \ref{eq:tconj}. Substituting in Eq.\:\ref{eq:thetaddot} yields the simple dimensionless differential equation,
\begin{equation}
\frac{\ddot{\theta}}{n^2} = \frac{1}{2\pi} \frac{\delta n}{n} \Big(\theta\Big) \label{eq:thetaddotgen}
\end{equation}
The problem therefore reduces to finding an expression for the fractional change in $n$ at conjunction, as a function of the conjunction angle $\theta$.

\subsection{Kicks at Conjunction} \label{sec:kick}
To estimate the kick $\delta n$ at conjunction, we estimate the change in energy $\delta E$ (we will take the test-particle limit $m \rightarrow 0$ at the end).
Assuming all fractional changes are small, Kepler's third law and $E=-GMm/(2a)$ imply
\begin{equation}
\frac{\delta n}{n} \approx -\frac{3}{2}\frac{\delta a}{a} \approx -\frac{3}{2}\frac{\delta E}{E},
\end{equation}
where $\delta a$ is the corresponding change to the semimajor axis (Fig.\:\ref{fig:encounter}).

\begin{figure}
    \centering \resizebox{\columnwidth}{!}{\includegraphics{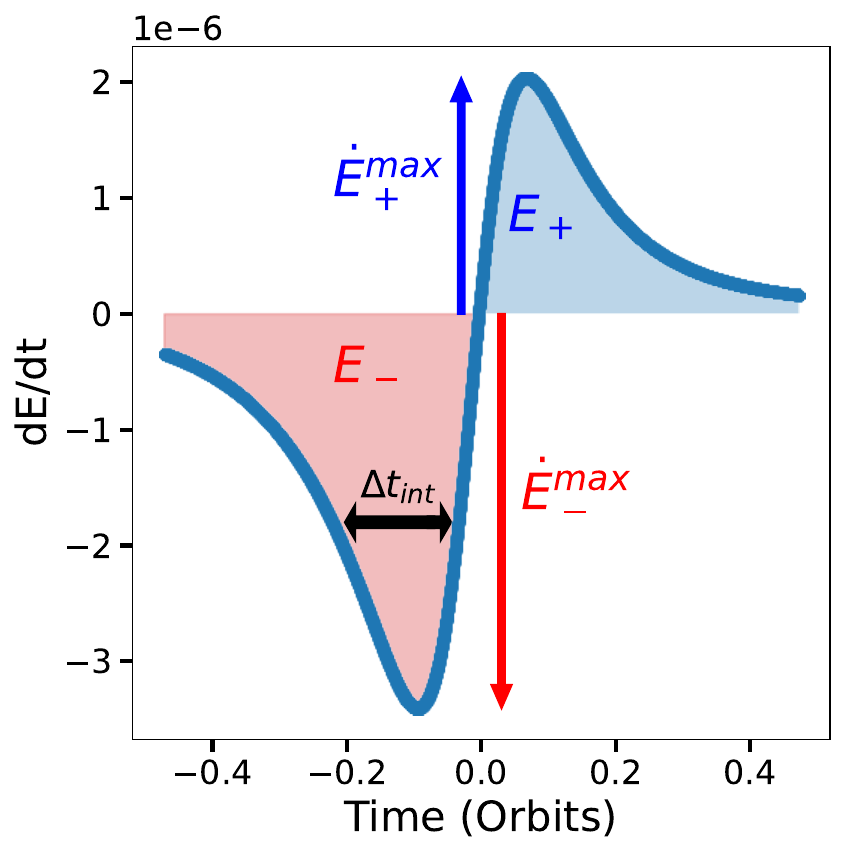}}
    \caption{Rate of change of the specific orbital energy of the test particle as a function of time near conjunction. Before conjunction, the test particle loses energy, with a maximum loss rate of $\dot{E}_-^\text{max}$, over a characteristic timescale $\Delta t_\text{int}$ (Eq.\:\ref{eq:dtint}). The total amount of specific energy lost is the area $E_-$ shaded red under the curve. The test particle then gains specific energy after the conjunction, with corresponding quantities in blue. Because the orbit is eccentric and the separation between the orbits is different before and after conjunction, the amounts of energy gained and lost are not equal and opposite.
    \label{fig:kick}}
\end{figure}

Prior to the inner body overtaking the outer, the outer test particle is being pulled backward opposite its velocity and thus loses orbital energy (Eq.\:\ref{eq:Edot}); after the close encounter it is pulled forward and gains orbital energy.
Both before and after conjunction, the rate of change of the energy (Eq.\:\ref{eq:Edot}) grows with the increasing force as the planets approach one another, but then falls to zero near conjunction when the force and velocity are perpendicular and the dot product vanishes (Fig.\:\ref{fig:kick}).

For circular orbits, the changes in energy (blue and red areas under the curve in Fig.\:\ref{fig:kick}) are equal and opposite to leading order in the perturbation (i.e., with the line integral taken over the unperturbed path); for eccentric orbits, this symmetry is broken, leading to a $\delta E = E_+ + E_-$ for which we can obtain the appropriate scalings.
Since the forces act on a timescale $\Delta t_{int}$ (Eq.\:\ref{eq:dtint}), the net change in energy should scale as
\begin{equation}
\delta E = \int \dot{E} dt \sim (\dot{E}_+^{max} + \dot{E}_-^{max}) \Delta t_\text{int}, \label{eq:deltaE}
\end{equation}
where we simply estimate the area under the curve before and after conjunction as rectangles with width $\Delta t_\text{int}$ and height given by their maximum of $\dot{E}$, labeled as blue and red arrows in Fig.\:\ref{fig:kick} (for exact expressions in terms of special functions, see \citealt{Namouni96}).

The maximum $\dot{E}$ occurs roughly when the azimuthal separation $y$ is comparable to the radial separation $\Delta a$ (Fig.\:\ref{fig:encounter}); this balances the vanishing dot product at conjunction where the force is maximum against the gravitational force decaying with increasing azimuthal separation.
In this case the cosine of the angle between the force and velocity vectors is of order unity and $\dot{E} \sim Fv$, and we can estimate the difference between $E_+^{max}$ and $E_-^{max}$ by considering the difference in radial separation before and after conjunction\footnote{In reality, the velocity and the angle between {\bf F} and {\bf v} also differ pre and post-conjunction, but these changes lead to the same scaling so we ignore them for simplicity.},
\begin{equation}
\dot{E}_+^{max} + \dot{E}_-^{max} \sim \Delta F v  \sim \Bigg(\frac{\partial F}{\partial r} \Delta r\Bigg) v ,
\end{equation}

Because $r = a(1-e\cos M) + \mathcal{O}(e^2)$, where $M=\lambda - \varpi$ is the mean anomaly, and since $M=\theta$ at conjunction (Eq.\:\ref{eq:theta}), we have 
\begin{eqnarray}
\Delta r &\approx& ae\cos (\theta + \Delta M) - ae\cos (\theta - \Delta M) \nonumber \\
&=& 2ae \sin\Delta M \sin \theta, \label{eq:deltar}
\end{eqnarray}
where $\Delta M \equiv n \Delta t_{int}$ is how far from conjunction the maxima in Fig.\:\ref{fig:kick} occur, and is of order unity given Eq.\:\ref{eq:dtint}.

Combining Eqns \ref{eq:deltaE}-\ref{eq:deltar}, and dropping factors of order unity, we find that the kicks to the mean motion scale as
\begin{equation}
\frac{\delta n}{n} \sim \frac{\mu \tilde{e}}{e_c^2} \sin \theta, \label{eq:nkick}
\end{equation}
where $\mu \equiv m_p/M_\star$ is again the planet-star mass ratio, $\tilde{e}$ is the eccentricity normalized to the orbit-crossing value (Eq.\:\ref{eq:etilde}), and we have used $\Delta a/a = e_c$.
We note that when conjunction happens at pericenter or apocenter and $\sin \theta = 0$, everything is symmetrical and $\delta n=0$.

\subsection{Pendulum Dynamics} \label{sec:pendulum}

Plugging in the expression for the kicks at conjunction (Eq.\:\ref{eq:nkick}) into our differential equation for $\theta$ (Eq.\:\ref{eq:thetaddot}),
\begin{equation}
\ddot{\theta} \sim A_1^2 \frac{\mu \tilde{e}}{e_c^2} n^2 \sin \theta \equiv \omega^2 \sin \theta, \label{eq:thetaddot1storder}
\end{equation}
where we have introduced a (squared) order-unity coefficient $A_1^2$ to make up for our simple scaling arguments (see Table \ref{tab:Aq}).

This is the differential equation of a simple pendulum, with its equilibrium at $\theta = \pi$ as expected from Sec.\:\ref{sec:qualitative} \footnote{In order to obtain a more familiar analogy to a pendulum, many authors define the conjunction angle $\theta'$ as measured from where the orbits are farthest apart so $\theta = \theta' + \pi$ \citep[e.g.,][]{Namouni96, Murray99}. One can check by substitution that this flips the sign of Eq.\:\ref{eq:thetaddot1storder}, yielding an equilibrium at $\theta' = 0$ like a typical pendulum. However, we will see that this latter choice leads to several undesirable features for higher order MMRs, so we retain our own (Sec.\:\ref{sec:higher}).}, and an angular frequency for small oscillations of
\begin{equation}
\omega = A_1\frac{\sqrt{\mu \tilde{e}}}{e_c} n. \label{eq:omega}
\end{equation}

The pendulum has a conserved energy

\begin{equation}
H = \frac{\dot{\theta}^2}{2} + \omega^2 \cos \theta, \label{eq:H}
\end{equation}
which we can use to calculate the width of the resonance. 
The separatrix corresponds to the pendulum trajectory that is nudged from rest ($\dot{\theta} = n - n_\text{res} = 0$, see Eq.\:\ref{eq:thetadot}) at the unstable fixed point at $\theta= 0$ (so plugging into Eq.\:\ref{eq:H}, $H_\text{sep} = \omega^2$), and oscillates back and forth with maximum amplitude.
The MMR width corresponds to the maximum $n - n_\text{res}$ attained on this separatrix trajectory (Fig.\:\ref{fig:phasespace}), which in turn corresponds to the maximum $\dot{\theta}$ reached through Eq.\:\ref{eq:thetadot}. 
We can solve the maximum $\dot{\theta}$ by equating the conserved Hamiltonian at this moment when the pendulum is swinging fastest through the stable equilibrium at $\theta = \pi$ (so $H_\text{sep} = \dot{\theta}_{max}^2/2 - \omega^2$) to the initial value ($H_\text{sep} = \omega^2$), yielding $\dot{\theta}_{max} = 2\omega$.
Using Eq.\:\ref{eq:thetadot}, we have
\begin{equation}
\Bigg|\frac{n-n_\text{res}}{n}\Bigg|_{max} \approx |\Delta_\text{max}| = \frac{3}{2}e_c \frac{\dot{\theta}_{max}}{n} = 3 A_1 \sqrt{\mu \tilde{e}}, \label{eq:widthq1}
\end{equation}
where we are only interested in absolute values given that the maximum deviation from resonance happens symmetrically above and below the resonant value in the pendulum approximation.
This result matches Eq.\:\ref{eq:width} for a first-order MMR ($q=1$).

\subsection{A Universal Model for MMRs} \label{sec:normalizedMMRs}
\begin{figure}
    \centering \resizebox{\columnwidth}{!}{\includegraphics{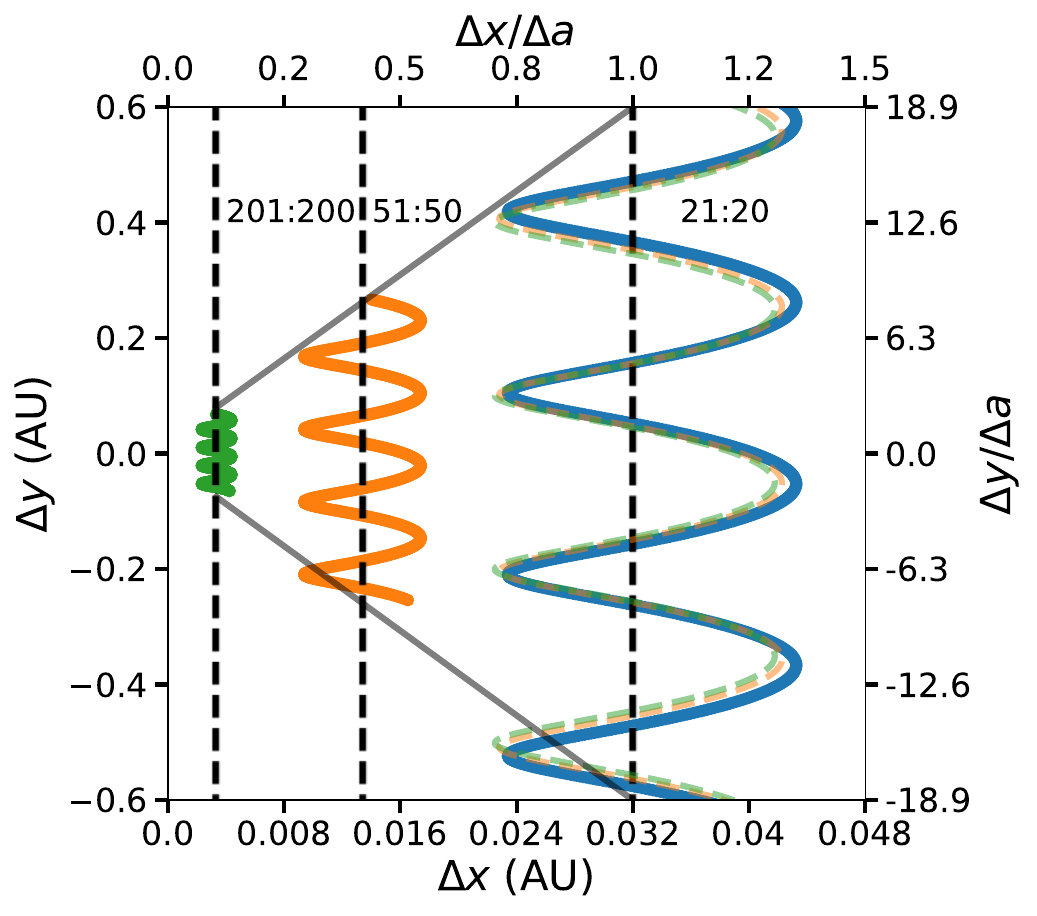}}
    \caption{Unperturbed motion of an eccentric test particle in Hill's coordinates for first-order MMRs corresponding to a wide range of separations (green, orange and blue solid curves). If we scale each curve's coordinates by the separation $\Delta a$ (dashed lines, top and right axes), the three curve's approximately overlap. 
    This shows that different MMRs of the same order are scaled versions of one another, and explains qualitatively why all first-order MMRs should share the same order-unity prefactor when integrating the effect of the massive planet along unperturbed paths in these normalized coordinates.
    \label{fig:normalizedMMRs}}
\end{figure}

Throughout this paper (and in much of celestial mechanics), calculations are done to first order in the perturbation, i.e., considering the total change in energy by integrating along \textit{unperturbed} elliptical trajectories.
In reality, the actual orbits deviate from these Keplerian paths near conjunction, but because we assume the perturbations remain small compared to the dominant gravity of the central body, this is a good approximation.

To avoid orbit-crossing, the test particle's radial excursions $ae$ need to be smaller than the separation $\Delta a$. 
For closely spaced orbits ($\Delta a/a \ll 1$), the eccentricities are thus always small.
In this Hill limit, one can approximate the unperturbed Keplerian ellipses to leading order in the eccentricities, yielding the so-called guiding center approximation \citep[e.g., Chapter 2.6 of][]{Murray99}.

The simplest approximation to a near-circular orbit is a body moving on a circle around the central body with the same mean motion $n$.
One obtains the leading $\mathcal{O}(e)$ correction to this motion by fixing an ellipse to the guiding center that rotates along with it.
To account for the radial variations, the ellipse's minor axis is $ae$, while the major axis (always pointing along the azimuthal direction) is $2ae$ to correct for the larger changes in the along-orbit direction.
The position of the body is then obtained by considering the superposition of the guiding center rotating at a rate $n$, with the epicyclic motion along the affixed ellipse at the same rate---in the opposite direction so that the orbital rate is faster at pericenter and slower at apocenter, see Fig.\:2.8 in \citep{Murray99}.

If we then write down the test particle's $x$ (radial) and $y$ (azimuthal) positions in a frame rotating with the inner planet's mean motion, we have

\begin{align}
x = ae \cos(nt) + \Delta a \\
y = -2ae \sin(nt) - \frac{3}{2} \Delta a nt
\end{align}
where we have used that for small separations, $\Delta n / n = \frac{3}{2} \Delta a / a$.
While this is a more complicated conceptual picture, this has replaced a time-varying orbital rate with a prescription that only has a single angular rate $n$.
More importantly for our purposes, the path is now linear in the eccentricities. 
If we nondimensionalize distances by the separation $\Delta a$ and time such that $n=1$, we have
\begin{align}
\tilde{x} = \frac{x}{\Delta a} = \tilde{e} \cos(t) + 1 \\
\tilde{y} = \frac{y}{\Delta a} = -2\tilde{e} \sin(t) - \frac{3}{2}t,
\end{align}
where $\tilde{e} = e/e_c$ is the same normalized eccentricity used above.

This means that while in the general problem, each first-order MMR has its own $\Delta a$ and $e_c$, this transformation results in the same normalized, unperturbed paths for all MMRs.
We show this graphically in Fig.\:\ref{fig:normalizedMMRs}, where we consider test particles at three different first-order MMRs (solid blue, orange and green lines), and show that they all yield approximately the same motion when normalized by the separation $\Delta a$ (dashed lines).
This implies that while we had to introduce a dimensionless order-unity coefficient $A_1$ to the scaling in Eq.\:\ref{eq:thetaddot1storder}, it will be universal to all closely spaced first-order MMRs.
The same argument holds for higher order $p$:$p-q$ MMRs with $q>1$.
The proportionality constant $A_q$ will be different, but it will apply to all MMRs of the same order $q$ (Table \ref{tab:Aq}).

\subsection{Limits of the Approximation} \label{sec:limits}

We have shown above that in the limit of close separations, one can extract the appropriate scaling of the coefficient for the cosine potential governing the dynamics (Eq.\:\ref{eq:H}). 
Our approximations deteriorate as $\Delta P/P$ approaches and exceeds unity and the unperturbed paths deviate from those in the Hill Limit (Sec.\:\ref{sec:normalizedMMRs}), i.e., near and beyond the 2:1 MMR, when the deviations from the universal prefactors $A_q$ (Table \ref{tab:Aq}) become substantial.

However, even at close separations, we arrived at a pendulum model by treating the eccentricities as constant.
This is a good approximation at high eccentricities, when the oscillations around the initial value can be ignored, but breaks down in the limit of circular orbits, where Eqs.\:\ref{eq:omega} and \ref{eq:widthq1} would predict a vanishing width and libration frequency for the MMR.
%In reality, in our notation, the test particle gets a kick to its normalized eccentricity vector $\delta \tilde{e}_\text{conj}$ of magnitude \citep{Duncan89}
%\begin{equation}
%\delta \tilde{e}_\text{conj} = \frac{\pi A_1^2 \mu}{e_c^3}.
%\end{equation}

%\memodt{Kick above is right, but these build up to larger ecc, so it's not the right threshold. Would help tu understand what the right mean eccentricity is for higher order MMRs beyond first? Fix this section later}

A more complete treatment would track the coupled evolution of the period ratio and the eccentricity.
This more complete treament leads to a description of the dynamics in terms of what \citet{HenrardLemaitre1983} dub the ``second fundamental model of resonance" (the first being the pendulum).
In this more complicated model, the equilibrium period ratio is no longer necessarily exactly at the corresponding integer ratio, and the proximity to resonance is a function of both the period ratio and the eccentricity \citep{HenrardLemaitre1983, Deck13, Batygin13, Hadden19}.

The failure of our above expressions is that one has effectively chosen an initial condition with an eccentricity that is not representative of the values of the eccentricity over the many conjunctions in the resonant cycle.
One can show that in this more sophisticated model, circular orbits initialized near period ratios corresponding to first-order MMRs oscillate around (normalized) equilibrium eccentricities $\tilde{e}_\star$ \citep[e.g.,][]{Ferraz07, Deck13}
\begin{equation}
\tilde{e}_\star \sim \Big(\frac{\mu}{e_c^2}\Big)^{1/3}.
\end{equation}

With this estimate for the average eccentricity in hand, one can then still use the pendulum model to obtain the appropriate scalings for first-order $p$:$p-1$ MMRs in this low-eccentricity limit from Eqs.\:\ref{eq:omega} and \ref{eq:widthq1}
\begin{eqnarray}
\text{Low-e Limit $j$:$j-1$ MMR:\:\:\:\:} \tilde{e} &\ll& \Big(\frac{\mu}{e_c^2}\Big)^{1/3} \nonumber \\
\omega &\sim& \frac{\mu^{2/3}}{e_c^{4/3}}n \nonumber \\
\Delta_\text{max} &\sim& \frac{\mu^{2/3}}{e_c^{1/3}}, \label{eq:lowecc}
\end{eqnarray}
which become independent of the eccentricity.

We defer a discussion of the limitations of the close-spacing approximation and the particuliarities of the 2:1 MMR to Sec.\:\ref{sec:21}, after we generalize our results to higher order MMRs and to two massive planets.

%This can be straightforwardly applied to, e.g., obtain the chaos criterion of \cite{Wisdom80} at which adjacent first-order MMRs overlap for initially circular orbits. The spacing $\Delta_\text{adj}$ between neighboring first-order MMRs at close separations is
%\begin{equation}
%\Delta_\text{adj} \sim \frac{\Delta P}{P} \sim \Bigg(\frac{p+1}{p} - \frac{p}{p-1}\Bigg) \sim e_c^2,
%\end{equation}
%so setting $\Delta_\text{max} = \Delta_\text{adj}$ and using Eq.\:\ref{eq:lowecc} yields the critical value of $e_c$ (i.e., the critical orbital spacing) below which the dynamics are expected to be chaotic \citep{Wisdom80, Deck13},
%\begin{equation}
%\Bigg(\frac{\Delta a}{a}\Bigg)_\text{crit} = e_c^\text{crit} \sim \mu^{2/7}.
%\end{equation}

%For the higher order MMRs discussed next, the dynamics of initially circular orbits tend to be more complicated because (unlike first-order MMRs), this initial condition can be a stable or unstable equilibrium depending on the initial period ratio and other parameters \citep{Ferraz07}. 
%However, a stronger scaling of the resulting eccentricities with the (small) mass ratio $\mu$ ($\mu^{1/2}$ for $k=2$, $\mu$ for $k=3$) implies that the low-eccentricity limit for these higher-order MMRs corresponds to much smaller eccentricities, and the pendulum approximation we make in this paper typically works well for non-zero eccentricities.
%For a detailed summary of the dynamics near higher order MMRs in this second fundamental model of resonance, see Appendix D of \cite{Ferraz07}.

\section{Higher-Order Resonances} \label{sec:higher}

We are now in a position to understand the behavior of higher-order MMRs, with period rations near a $p:p-q$ commensurability for $q>1$, e.g., the second-order 5:3 MMR with $q=2$. 
The classical result from celestial mechanics, where one isolates resonant harmonics from a Fourier expansion of the gravitational interaction potential between the two planets, is that the ``strengths" of MMRs scale as $e^q$ (i.e., the coefficient of the resonant cosine term in the disturbing function, Eq.\:\ref{Eres}).
We found above from physical grounds that this coefficient ($\omega^2$ in Eq.\:\ref{eq:H}) indeed scales linearly with eccentricity for first-order MMRs with $q=1$ (see Eq.\:\ref{eq:omega}).

\subsection{Conjunction Effects Cancel to Yield Weak High-Order MMRs}
Following our above analysis, we can now understand this behavior quite simply. For the first-order MMRs analyzed above, the inner planet executes one more orbit than the inner planet per cycle, resulting in a single conjunction.
For, e.g., the second-order 5:3 MMR, the inner planet would make five orbits in the time the outer body does three, which means the inner planet had to overtake twice. 
The order of the resonance $q$ therefore corresponds to the number of conjunctions before the cycle repeats.
%For a $p:p-q$ MMR, the inner planet orbits $p$ times in the time it takes the outer planet to orbit $p-q$ times, and then the cycle repeats.
%This means that the inner planet overtakes $q$ times each cycle.%, so each cycle lasts $q t_{conj}$.

In Fig.\:\ref{fig:second}, we showed a plot analogous to the left panel of Fig.\:\ref{fig:qualitative}, now considering a second-order 5:3 MMR. 
Because the planetary perturbations are small and the planets orbit at approximately constant rates, the two conjunctions in the cycle must be approximately equally spaced in time and occur on opposite sides of the orbit (see discussion in Sec.\:\ref{sec:higherintro}).

The dynamics of each of these conjunctions is identical to our discussion above, but the conjunction locations for the 5:3 (Fig.\:\ref{fig:second}) imply that they will have approximately equal and opposite effects. 
Because the orbits are becoming more widely spaced at conjunction 1, the test particle speeds up, while at conjunction 2 when the orbits are approaching one another, the test particle slows down (Sec.\:\ref{sec:qualitative}).

\begin{figure}
    \centering \resizebox{\columnwidth}{!}{\includegraphics{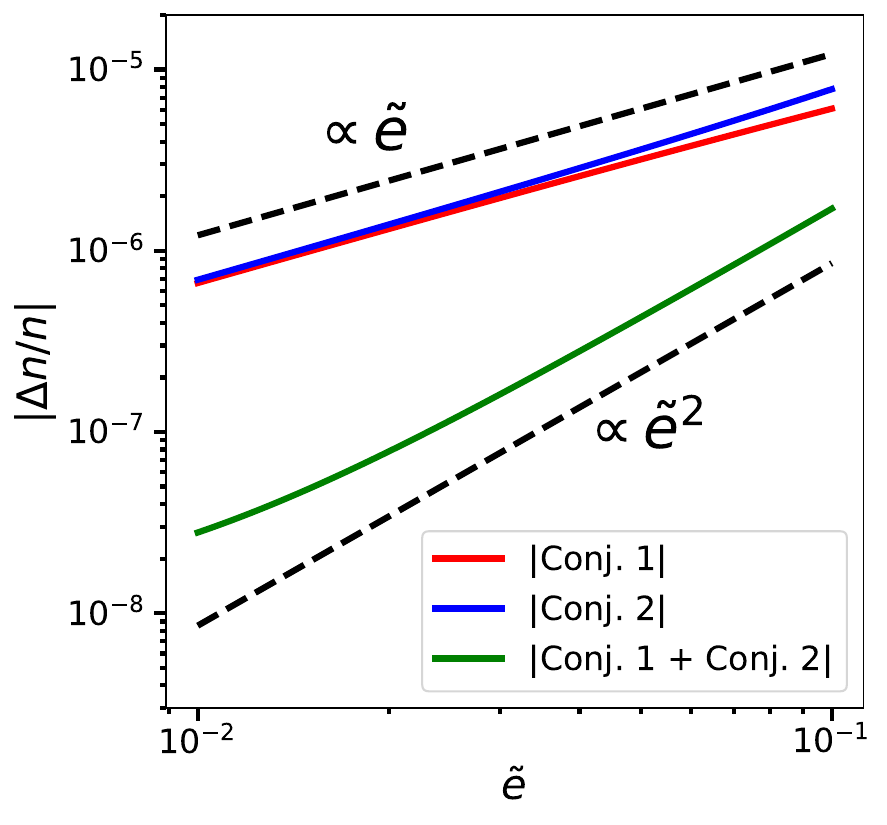}}
    \caption{A $q$th order MMR has $q$ conjunctions per cycle. The plot shows (for a second-order MMR) that while the individual kicks to the mean motion at conjunctions always scale linearly with eccentricity (red and blue lines), their opposite signs cause their total effect to cancel to leading order, explaining the weaker eccentricity scaling of higher order MMRs (green line). Specifically, the plot shows the fractional changes in the mean motion $\Delta n/n$ caused by conjunctions for a 5:3 MMR, as a function of the initial orbital eccentricity $\tilde{e}$ (normalized to the value at which the orbits would cross). The red and blue lines show the linear eccentricity-scaling for the kicks to the mean motion from each of the two individual conjunctions per cycle. The green line shows the magnitude of the sum of the two kicks per cycle, which cancel to leading order in $\tilde{e}$ due to their opposite signs. This results in a $\tilde{e}^2$ scaling as appropriate for second-order MMRs (green line). All simulations initialize an Earth-mass planet at 1 AU around a solar-mass star, with an exterior, eccentric test particle ($\tilde{e}$ varied along x-axis) with a period ratio of 5:3. The angles are initialized so that conjunctions occur as shown in Fig.\:\ref{fig:second}.
    \label{fig:scaling2}}
\end{figure}

In Fig.\:\ref{fig:scaling2}, we show that the magnitude of the kick to the mean motion scales linearly with eccentricity for each individual conjunction (red and blue lines) as expected from our analysis above.
However, if we account for their sign and add up the two kicks to obtain the net change in the mean motion each resonant cycle, we obtain the green line, which scales as $e^2$ as appropriate for a second-order MMR.
For third-order MMRs, the approximate 120$^\circ$ symmetry causes the mean motion kick to cancel at both first and second order, resulting in a $e^3$ scaling, and so on for higher order MMRs.

\subsection{Power Series for the Kicks at Conjunction} \label{sec:powerseries}

Our simple arguments above yielded the leading-order scaling for the kicks at conjunction.
In reality, there are also higher-order contributions from all the harmonics of $\theta$ in the periodic gravitational potential between the two planets, which are worked out in detail by \cite{Namouni96}.
In our notation, we can write the fractional kicks to the mean motion to leading order in $\mu$ as\footnote{This differs from the notation in \cite{Namouni96} both through a simple normalization to simplify our later expressions, and by defining our conjunction angle relative to the longitude of minimum (rather than maximum) separation. While the corresponding coefficients $W_q^{q,0}$ in \cite{Namouni96} alternate sign according to whether $q$ is even or odd, our choice leads to coefficients $A_q^2 = \frac{4}{3}\frac{|W_q^{q,0}|}{2\pi}$ that are always positive and lead to simpler expressions for widths, frequencies and equilibria.}
\begin{equation}
\frac{\delta n}{n} \approx 2\pi \frac{\mu}{e_c^2} \sum_{j=1}^\infty j A_j^2 \tilde{e}^j \sin (j \theta), 
\label{eq:namkick}
\end{equation}
%\begin{eqnarray}
%\frac{\delta n}{n} &=& -\frac{\partial W}{\partial \theta} \approx 2\pi \frac{\mu}{e_c^2} \sum_{j=1}^\infty j A_j^2 \tilde{e}^j \sin (j \theta), \nonumber \\
%W &=& \frac{4}{3}\frac{\mu}{e_c^2}\sum_{j=0}^\infty W_j \cos (j\theta) \nonumber \\
%W_j &=&\sum_{k=0}^\infty W_j^{j+2k,0}e^{j+2k}
%\label{eq:namkick}
%\end{eqnarray}
where the $A_j$ are the dimensionless coefficients in Table \ref{tab:Aq}.
Since we assume that the orbits do not cross, $\tilde{e} < 1$ and the terms in the sum are successively smaller\footnote{We also neglect additional corrections to the coefficients of each $\sin(j\theta)$ term in Eq.\:\ref{eq:namkick} of order $\tilde{e}^{j+2}$ that do not affect our arguments below.}. 
%zeroth order in $e$ and second order in $\mu$ contribution to eq namkick : \frac{\pi^2 A_1^4 \mu^2}{e_c^5}

We can see that for the 5:3 MMR example above, where the conjunctions happen on opposite sides of the orbit, the leading term in the series cancels, since one kick is $\propto \sin \theta$, while the other is $\propto \sin (\theta + \pi) = -\sin{\theta}$.
By contrast, the $q=2$ terms add up, since $\sin 2(\theta + \pi) = \sin 2\theta$.
In general, for a $q$th order MMR, the first $q-1$ terms in the series will cancel, so the leading order term in the eccentricities scales as $e^q$.
Given that our focus for this paper is on physical intuition, we defer a more mathematical analysis of this general result to a follow-up paper\footnote{We further note that our simple argument assumes a single eccentricity for all the conjunctions in the cycle, and that these conjunctions are equally spaced in longitude. For high enough order MMRs (where the surviving leading-order $e^q$ term is small enough), these small corrections would have to also be taken into account.}.

The key insight is that \textit{all} MMRs can be understood as due to the effects of repeated conjunctions in the compact limit; however, higher order MMRs are weakened by the cancelling effects of symmetrically spaced conjunctions.
It provides a simple, physical answer for why observational features are typically only prominent for low-order MMRs, and why the dense set of higher-order MMRs can be ignored.

\subsection{MMR Widths and Libration Frequencies} \label{sec:widthsgen}

If one traces the arguments above for first-order MMRs (Secs.\:\ref{sec:conjunctionangle}-\ref{sec:kick}), they apply equally well to a $q$th order MMR, leading to the same differential equation for $\theta$ (Eq.\:\ref{eq:thetaddotgen}).
In particular, there is still a kick occurring every $t_\text{conj}$ (Eq.\:\ref{eq:tconj}); the only difference is that the terms in the sum in Eq.\:\ref{eq:namkick} with $j<q$ will cancel.
Substituting the leading order term in Eq.\:\ref{eq:namkick} with $j=q$ into Eq.\:\ref{eq:thetaddotgen}, we have 
\begin{equation}
\frac{\ddot{\theta}}{n^2} = q A_q^2 \frac{\mu \tilde{e}^q}{e_c^2}  \sin (q \theta)
\end{equation}

We now switch variables to the traditional resonant angle $\phi = q\theta$, yielding
\begin{equation}
\frac{\ddot{\phi}}{n^2} = q\frac{\ddot{\theta}}{n^2} = q^2 A_q^2 \frac{\mu \tilde{e}^q}{e_c^2} \sin \phi \equiv \omega^2 \sin \phi, \label{eq:phiddot}
\end{equation}
where the frequency of small oscillations is
\begin{equation}
\frac{\omega}{n} = q A_q \frac{\sqrt{\mu \tilde{e}^q}}{e_c}, \label{eq:omegagen}    
\end{equation}
and the pendulum has a conserved energy
\begin{equation}
H = \frac{\dot{\phi}^2}{2} + \omega^2 \cos \phi. \label{eq:Hgen}
\end{equation}

This is the same functional form as Eq.\:\ref{eq:H}, so we similarly have that $\dot{\phi}_\text{max} = 2\omega$ on the separatrix trajectory.
However, one must be careful in converting back to $\dot{\theta}_\text{max} = \dot{\phi}_\text{max}/k$ to obtain the MMR width from Eq.\:\ref{eq:thetadot}, yielding
\begin{equation}
\Bigg(\frac{n-n_\text{res}}{n}\Bigg)_{max} = \frac{3}{2}e_c \frac{\dot{\phi}_{max}}{qn} = 3 A_q \sqrt{\mu \tilde{e}^q}. \label{eq:widthgen}
\end{equation}

In the pendulum approximation we explore in this paper, the introduction of the resonant angle $\phi$ thus yields identical dynamics for MMRs of all orders (albeit with different widths and frequencies).
This implies that the stable equilibrium is always at $\phi = \pi$ as we found before\footnote{This is again due to our choice to measure the conjunction angle $\theta$ relative to the location where the orbits are closest together.
The alternate choice of defining $\theta$ relative to where they are furthest apart leads to the stable equilibrium flipping from 0 to $\pi$ for even and odd order MMRs, respectively \citep{Namouni96, Murray99}.}.
This implies that there is one stable conjunction angle at $\theta_\text{eq} = \pi/q$, with the remainder spaced symmetrically $2\pi/q$ radians apart, e.g., at $\pi/2$ and $3\pi/2$ for second-order MMRs, at $\pi/3$, $\pi$ and $5\pi/3$ for third-order MMRs, etc. (see Fig.\:\ref{fig:folding}).

\section{An approximate mapping from two massive planets on compact, eccentric orbits to the CR3BP} \label{sec:general}

We finally explore why the simpler case of a test particle perturbed by a massive planet on a fixed circular orbit (the co-planar, circular restricted 3-body problem) provides a useful model for the dynamics of two massive planets on eccentric orbits in the compact limit.

\cite{Hadden19} shows that each of the simplifications below applies not only to the dynamics at close separations in the Hill limit, but also in the classical celestial mechanics treatment where one isolates resonant harmonics from the disturbing function expansion. 
The latter approach through canonical perturbation theory yields more general expressions for conserved quantities, but they are harder to interpret. 

A central motivation for this paper is that the fact that the meaning of these various quantities can be physically understood in the Hill limit provides meaningful labels and ways to talk about these otherwise more opaque variables in the celestial mechanics approach. 
We discuss each of them in detail below.

The key result is that near conjunction in the Hill limit, the dynamics separate cleanly into two independent degrees of freedom: one for the secondaries' center of mass, and another for their relative separation \citep{Hill1878, Henon69, Henon86}.
We now explore the consequences of this fact together with the physical reasons for this behavior.
In this section we label the inner and outer planets with subscripts of 1 and 2, respectively.

\subsection{The Relative Eccentricity} \label{sec:erel}

\begin{figure}
    \centering \resizebox{\columnwidth}{!}{\includegraphics{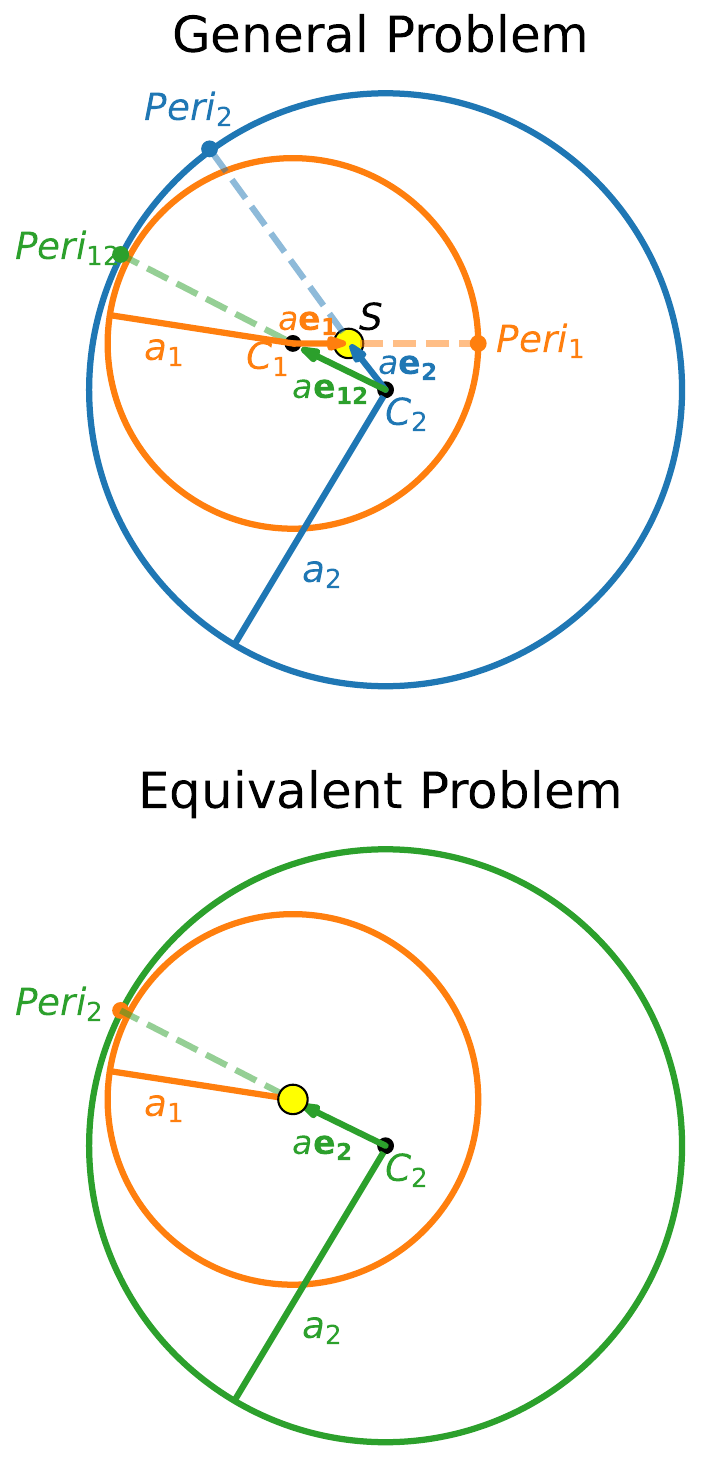}}
    \caption{To leading order in the eccentricity, eccentric orbits are equivalent to offset circles (the inner orbit in orange becomes a circle of radius $a_1$ centered at $C_1$, while the outer blue orbit is centered at $C_2$). This makes it possible to map the encounter geometry in the general problem of two eccentric orbits to one where the inner planet is on a circular orbit around the Sun (bottom panel), and the outer planet has eccentricity vector $\mathbf{e_2} = \mathbf{e_{12}}$.
    \label{fig:equiv}}
\end{figure}

We first ask which combination of eccentricities should enter into the equations for the MMR widths and frequencies.
The dependence of the dynamics on the relative separation implies that the key combination is the ``relative eccentricity" vector $\mathbf{e_{12}} = \mathbf{e_{2}} - \mathbf{e_{1}}$ (Eq.\:\ref{eq:erel})
We can obtain a geometric intuition for this result in the following way.

In the compact limit, the eccentricities need to be $\ll 1$ in order not to cross.
To $\mathcal{O}(e)$, an elliptical orbit can be approximated as a circle of radius $a_i$ whose center is offset by $a_ie_i$ away from pericenter, e.g., the center $C_1$ of the inner orbit in Fig.\:\ref{fig:equiv} in orange is offset away from the star $S$ by a distance $a_1e_1$.
In the compact limit, $a_1 \approx a_2$, so we can simplify things by calculating these offsets using a common semimajor axis, which we define as the center-of-mass value $a$ (this choice will simplify the interpretation in the next subsection)
\begin{equation}
a = \frac{m_1 a_1 + m_2 a_2}{m_1+m_2}. \label{eq:a}
\end{equation}
We can then see from Fig.\:\ref{fig:equiv} that the geometry of two closely packed, eccentric orbits around a star $S$ is then same as a problem where the star is instead located at $C_1$, the inner planet is on a \textit{circular} orbit with semimajor axis $a_1$, and the outer planet is on an eccentric orbit of semimajor axis $a_2$ around $C_1$ with $\bf{e} = \bf{e_{12}}$. 
This provides a mapping from the encounter geometry of two eccentric orbits back to the simpler case analyzed above where the inner orbit is circular (Sec.\:\ref{sec:scaling}).

It also provides a simple picture for why there is only one combination of the eccentricities that matters.
If one increases the two orbits' eccentricities by the same amount and in the same direction (i.e., with the same pericenter direction) in Fig.\:\ref{fig:equiv}, one simply offsets the circles by the same amount, and the relative separations between the orbits remain unchanged.
Only by changing the relative eccentricity vector do the geometry and strength/asymmetry of interactions at conjunction differ.
The dynamics (e.g., the MMR widths and frequencies) should therefore only depend on the relative eccentricity $\bf{e_{12}}$.

We note that this simple geometric argument applies to the paths swept out by the planets, but does not account for the varying orbital rates along those eccentric orbits.
One might worry that this could influence the net energy exchange at conjunction.
However, the same linearity in the eccentricities that we exploited to simplify the geometry above applies to the differences in the orbital rates, so that the mapping from the general case of two eccentric orbits to the circular case is preserved.
We show this explicitly in Appendix \ref{sec:erelHill}.

%\begin{equation}
%    \Delta \delta n/n  \sim \mu \tilde{e}/e_c^2 \sin \phi + \mu (\tilde{e} + O(e))/e_c^2 \sin (-\phi + O(e)) \sim O(e^2)
%\end{equation}

\subsection{The Resonant Angle} \label{sec:collapse}

An important technical difficulty in the traditional celestial mechanics approach to the general case of two massive planets on co-planar orbits is the need to consider several separate resonance angles for a given MMR.
For example, for the 5:3 MMR one needs to consider three terms
\begin{eqnarray}
&& ^{5:3}C_1 e_1^2 \cos(5\lambda_2 - 3 \lambda_1 - 2\varpi_1) \nonumber \\
&& ^{5:3}C_2 e_1e_2 \cos(5\lambda_2 - 3 \lambda_1 - \varpi_1 - \varpi_2) \nonumber \\
&& ^{5:3}C_3 e_2^2 \cos(5\lambda_2 - 3 \lambda_1 - 2\varpi_2), \label{eq:cosi}
\end{eqnarray}
where the $^{5:3}C_i$ are numerical coefficients in the disturbing function expansion \citep[e.g.,][]{Murray99}, whose combinatorics require that each cosine term scale as $e_1^p e_2^q$, where $p$ and $q$ are the coefficients of the $\varpi_i$ in the resonance angle (there is thus only one resonance angle to consider in the CR3BP we analyzed above, since $e_1=0$ implies that any cosine terms involving $\varpi_1$ vanish).

The fact that the pericenter is fixed in the two-body problem implies that the variations in the $\varpi_i$ (due to the other planet) are always much slower than the evolution of the $\lambda_i$ (due principally to the star). 
If the period ratio is close to 5:3, all three of these angles will then vary slowly and have to be accounted for simultaneously.
For a $q$th order resonance, there are in general $q+1$ resonant angles to consider, due to the requirement that the coefficients in any valid resonant angle must sum to zero (d'Alembert relation).\footnote{Physically admissible $\phi_i$ need to be invariant to an arbitrary choice of reference direction (i.e., $\lambda_i' = \lambda_i + \beta$ and $\varpi_i' = \varpi_i + \beta$). This will only be true if the sum of the coefficients in the resonance angle is zero, e.g., 5-3-2=0.} 
This presents a significant technical obstacle to analytical progress, and a conceptual barrier to a pendulum interpretation involving a single cosine potential.
Nevertheless, Fig.\:\ref{fig:equiv} again provides valuable physical intuition.

As we argued in the previous subsection, the relative geometry of the two orbits does not depend on the $e_i$ and $\varpi_i$ independently, but rather only on the relative eccentricity and its corresponding pericenter, $e_{12}$ and $\varpi_{12}$.
\cite{Hadden19} shows that this limiting behavior is encoded in the disturbing function coefficients for all closely spaced MMRs (i.e., the $^{5:3}C_i$ in the case of the 5:3), such that the $k+1$ MMR terms can be collapsed into a single one of the form 
\begin{equation}
s_2(\tilde{e}) \cos (5\lambda_2 - 3 \lambda_1 - 2\varpi_{12}), \label{eq:rescombined}
\end{equation}
where $s_2$ is a function of the normalized eccentricity universal to all second order MMRs.
For general orders $q$, \cite{Hadden18} additionally show that, to leading order in $\tilde{e}$,
\begin{equation}
s_q(\tilde{e}) = \frac{3}{4} A_q^2 \tilde{e}^q \label{eq:sq}
\end{equation}
if expressed in our notation.
This ability to collapse resonant terms into a single cosine to leading order in the eccentricity was known to hold exactly for first-order MMRs \citep{Sessin84, Wisdom86}, but \cite{Hadden19} showed it also holds approximately for higher order MMRs, with the approximation improving with closer separations (as it must in the Hill limit).

This provides a mapping from the general co-planar case to the CR3BP with a single cosine potential, where one simply replaces the test particle's eccentricity and pericenter with $e_{12}$ and $\varpi_{12}$.

This also helps resolve otherwise opaque mathematical behaviors.
We argued in Sec.\:\ref{sec:pendulum} that for a first-order MMR, e.g. the 3:2, with an outer test particle, the stable equilibrium corresponds to $\phi_\text{eq} = 3 \lambda_2 - 2\lambda_1 - \varpi_2 = \pi$.
However, if instead we make the test particle the interior particle, the stable equilibrium would lie at $\phi = 3\lambda_2 - 2\lambda_1 - \varpi_1 = 0$.
These mathematical statements obfuscate the fact that in both cases, the stable equilibrium lies at the location where the two orbits are most widely separated (when the test particle is exterior, at its apocenter; when the test particle is interior, at its pericenter). 
Transforming these resonant angles to the above form yields $3\lambda_2 - 2\lambda_1 - \varpi_{12} = \pi$ in both cases, as the reader can readily verify by drawing figures equivalent to Fig.\:\ref{fig:equiv}.

Finally, we note that in the pendulum approximation we approximate both the eccentricities and pericenters as fixed.
In reality, the location of closest approach between the two orbits $\varpi_{12}$ slowly evolves with time.
Nevertheless, Fig.\:\ref{fig:qualitative} still gives the right physical picture; it just corresponds to a frame slowly rotating with the location of minimum separation to always keep it at the top of the figure.

\subsection{The Center-of-Mass Eccentricity}

The independent center-of-mass degree of freedom provides a remaining, linearly independent combination of eccentricities, which we show is always conserved in the Hill limit.
This bears a close analogy to the simpler two-body problem.

For example, one typically models the motion of a planet about its host star as a two-body problem, where their equal and opposite gravitational forces on one another maintain a fixed center of mass (COM), while their mutual separation sweeps out a Keplerian ellipse.
Of course, the COM is not actually fixed; on much longer timescales, it executes an orbit around the galaxy. 
The large ratio between the star-planet separation and the system's distance from the galactic center allows one to separate the system's coordinated COM motion around the galaxy by approximating the external forces on the star and planet as equal. 
%The star-planet dynamics would in principle incorporate the tidal external field left over after subtracting off the mean external field driving the COM motion (which would be negligible in this case).

Similarly, as two planets on closely spaced orbits approach conjunction, they feel approximately the same gravitational force from their central star. 
The two planets' mutual COM in the Hill problem therefore also follows a fixed two-body Keplerian orbit around the star with eccentricity vector

\begin{equation}
{\bf e_{COM}} = \frac{m_1 {\bf e_1} + m_2 {\bf e_2}}{m_1 + m_2}, \label{eq:ecom}
\end{equation}
as shown explicitly in Appendix \ref{sec:erelHill}.

The effects considered in this paper (Sec.\:\ref{sec:scaling}) of the equal and opposite forces the two planets exert on one another near conjunction therefore conserve the planets' COM, and affect only their relative motions.

One subtlety is that in the star-planet example, the two-body motion is a bound elliptical orbit, so the two bodies always remain close together as the system orbits the galaxy; in the Hill problem, the conjunction represents a hyperbolic encounter, after which the planets separate and get far from one another (at which point the force from the star will no longer be approximately equal for the two bodies).
However, we recall that the central simplification in the compact limit is that the planets follow effectively unperturbed two-body orbits during the time between conjunctions.
Thus, the fact that conjunctions conserve ${\bf e_{COM}}$, while unperturbed 2-body motion between conjunctions conserves ${\bf e_{1}}$ and ${\bf e_{2}}$ independently, implies that ${\bf e_{COM}}$ is always a conserved quantity in the Hill limit.

We can therefore imagine a Keplerian ellipse for the COM with semimajor axis $a$ and eccentricity vector ${\bf e_{COM}}$. We note that this orbit \textit{does not} trace out the path of the actual COM between the two planets\footnote{For example, at a time halfway between conjunctions, two equal-mass planets would be on opposite sides of the star and have their COM near $r=0$ rather than $r\approx a$.}; it instead traces out where the COM \textit{will be} when conjunction occurs, similar to $\theta$ only corresponding to the conjunction longitude at the time of conjunction (see discussion around Eq.\:\ref{eq:theta}).

Together with the dynamics of the relative eccentricity, this fully determines the evolution of ${\bf e_1}$ and ${\bf e_2}$. 
In the pendulum approximation of Sec.\:\ref{sec:scaling}, we treated the relative eccentricity as fixed, but the above applies to more detailed models of MMRs that account for the varying eccentricities \citep{Henrard82, Batygin13, Hadden19}.
In particular, $\bf{e_{12}}$ evolves while $\bf{e_{COM}}$ remains fixed (though, e.g., \citealt{Sessin84, Deck13} do not connect their related variables to these more geometrically interpretable ones.

\subsection{Mass dependence} \label{sec:mass}

In the two-body problem, the textbook reduction to an equivalent one-body problem involves identifying the sum of the two masses as the central mass driving the relative motion.
This is why, e.g., Kepler's third law depends specifically on the sum of the two masses, which is important to consider for comparable bodies e.g., stellar binaries.
For the same reason, the relative dynamics of the two secondaries in the Hill problem depends only on their total mass.
We can therefore apply the same results from the circular restricted problem in the general Hill problem, e.g., the MMR widths and frequencies of small oscillations, by identifying $\mu$ with the ratio of the total planetary mass to that of the star,

\begin{equation}
\mu = \frac{m_1 + m_2}{M_\star} \label{eq:mu}
\end{equation}

Analogously to the two-body problem, the mass ratio between the two bodies enters only in determining how the relative motion is distributed between the planets.
For example, the evolution of the \textit{ratio} of the mean motions in the right panel of Fig.\:\ref{fig:qualitative} would remain unchanged if we distributed part of the mass $m$ of the planet to the test particle; the mass ratio between the planets only determines the relative amplitude of variation in the individual planets' mean motions.
The mean motion variations would be equal and opposite for same-mass planets, but in the limit we originally adopted for Fig.\:\ref{fig:qualitative} where all the mass in the inner planet, the changes in the inner planet's mean motion vanish, and the changes are all reflected in the outer orbit.

This result is derived explicitly through detailed Hamiltonian perturbation theory for first-order MMRs by \cite{Deck13}, but the result is general to MMRs of all orders, and the physical intuition is simply that of the two-body problem.

\subsection{Period Ratios} \label{sec:ncom}

In the general problem of two massive planets there are two separate degrees of freedom associated with each planet's orbital period.
In the Hill limit, however, the separation of the center-of-mass degree of freedom separates the period behavior into a ``center-of-mass" and relative orbital period, just as it does for the eccentricities.
In particular, the relative mean motion $\Delta n = n_1 - n_2$ will oscillate, while the ``center-of-mass" mean motion acts as a steady, uniformly advancing clock
\begin{equation}
n_\text{COM} = \frac{m_1 n_1 + m_2 n_2}{m_1 + m_2}. \label{eq:ncom}
\end{equation}
Conservation of $n_\text{COM}$ implies that the individual mean motions oscillate out of phase with one another, in proportion to their masses\footnote{Another way to see this is that the (conserved) total energy $E = -GM_\star m_1/(2a_1) - GM_\star m_2/(2a_2) - U_\text{int}$, where $U_\text{int}$ is the gravitational interaction energy between the two planets. In the Hill limit, $U_\text{int} \approx 0$ away from conjunction, so $2E/(GM_\star) \approx -m_1/a_1 - m_2/a_2$ is conserved across a conjunction, implying $\frac{\delta n_1}{n_1} =-\frac{m_2}{m_1}\frac{a_1}{a_2}\frac{\delta n_2}{n_2}$, or $m_1 \delta n_1 \approx -m_2 \delta n_2$ for closely spaced orbits.}.

We define resonant reference values for the mean motions $n_1^\text{res}$ and $n_2^\text{res}$ by requiring that they have the appropriate integer ratio, and $(m_1 n_1^\text{res} + m_2 n_2^\text{res})/(m_1 + m_2) = n_\text{COM}$ to match initial conditions (Eq.\:\ref{eq:ncom}).

In the Hill limit, the dynamically relevant quantity is the deviation of the relative mean motion from resonance, $\Delta n - \Delta n_\text{res}$.
However, it is often more natural to discuss MMRs in terms of integer \textit{ratios} of mean motions rather than differences.
Since for typical planetary masses the widths of MMRs are small, one can expand $\Delta$ (Eq.\:\ref{eq:Delta}) to leading order in the small deviations from the resonant mean motions to obtain
\begin{equation}
\Delta \approx \Bigg(\frac{n_1 - n_1^\text{res}}{n_1^\text{res}}\Bigg) - \Bigg(\frac{n_2 - n_2^\text{res}}{n_2^\text{res}}\Bigg). \label{eq:Deltaapprox}
\end{equation}

We can generalize our expressions in previous sections for the case of an outer test particle in the CR3BP by replacing
\begin{equation}
\frac{n - n_\text{res}}{n} \rightarrow -\Delta,
\end{equation}
since in that case $n_1 = n_1^\text{res}$ always, and $n = n_2 \approx n_2^\text{res}$.

Given that the deviations in the numerator of Eq.\:\ref{eq:Deltaapprox} are already small, in the denominators we could also neglect the small differences between $n_1^\text{res}$ and $n_2^\text{res}$ from $n_\text{COM}$ in the compact limit, yielding
\begin{equation}
\Delta \approx \frac{\Delta n - \Delta n_\text{res}}{n_\text{COM}}.
\end{equation}
At our level of approximation, in all quantities that are already small, we are neglecting the additional differences between $n_1$, $n_2$ and $n_\text{COM}$, so it inconsistent to be pedantic about the choice of $n$ while dropping other terms at the same order. 

This leads to generalizations of Eqns \ref{eq:thetadot}, \ref{eq:thetaddot} and \ref{eq:namkick} of
\begin{eqnarray}
\frac{\dot{\theta}}{n_\text{COM}} &=& -\frac{2}{3} \frac{\Delta}{e_c} \nonumber \\
\frac{\ddot{\theta}}{n_\text{COM}^2} &=& -\frac{1}{2\pi} \delta \Delta (\theta) \nonumber \\
\delta \Delta &=& -2\pi \frac{\mu}{e_c^2} \sum_{j=1}^\infty j A_j^2 \tilde{e}_{12}^j \sin (j \theta). \label{eq:Deltaeqs} 
\end{eqnarray}
We note that when we plug in the third equation for the kick to $\Delta$ at conjunction into the second line, the negative signs cancel and we recover the same differential equation as we had before.

%Assuming small deviations from resonance $\delta n_i \equiv n_i - n_{i0}$ and expanding the mean motions to first order in the $\delta n_i$, one can show
%\begin{equation}
%\Delta = \frac{n_{COM}^2}{n_{10}n_{20}} \Bigg(\frac{\delta n}{n_{COM}}\Bigg) + \mathcal{O}\Bigg(\frac{\delta n}{n_{COM}}\Bigg)^2,
%\end{equation}
%which approaches the fractional kicks to the relative mean motion in the compact limit where $n_{10}\approx n_{20} \approx n_{COM}$.

%For general case, we can think of Lithwick $\Delta$ as either a function of $(n_1, n_2)$ or $(n_{COM}, n_{rel}$.
%\begin{equation}
%\delta \Delta (n_1, n_2) = \frac{\delta n_1}{n_1^{res}} - \frac{\delta n_2}{n_2^{res}}
%\end{equation}
%\begin{equation}
%\delta \Delta (n_{COM}, n_{rel}) = \Bigg(\frac{n_{COM}^2}{n_1^{res}n_2^{res}}\Bigg)\frac{\delta n_{rel}}{n_{rel}^{res}}
%\end{equation}

\section{Conclusion} \label{sec:conclusion}

In the compact limit, the dynamics near mean motion resonances can be understood as the result of successive gravitational kicks at conjunction.
For a $q$th order MMR of the form $p$:$p-q$, there are $q$ conjunctions before the cycle repeats, and the symmetrical distribution of the conjunction locations causes their effects to partially cancel (Sec.\:\ref{sec:higher}).
This provides a physical intuition for why high order MMRs are weak and can typically be ignored, while low-order MMRs have more prominent observational signatures.

In addition, we explored physical arguments for why the general case of two massive planets on closely spaced, eccentric, coplanar orbits can be mapped onto the much simpler PCR3BP (planar, circular restricted three-body problem, Sec.\:\ref{sec:general}), and why all MMRs of a given order can be treated on the same footing (Sec.\:\ref{sec:normalizedMMRs}). 
We summarize this correspondence, together with simple expressions for MMR widths and frequencies, in Sec.\:\ref{sec:summary}, and give simple scaling arguments for first-order MMRs in Sec.\:\ref{sec:scaling}.

We argue that these physically intuitive quantities in the Hill limit can provide a useful vocabulary for discussing the more complicated variable combinations obtained from canonical perturbation theory for arbitrary orbital separations \citep[e.g.,][]{Murray99, Ferraz07, Delisle12, Deck13, Batygin13, Petit17, Hadden19}.

\begin{acknowledgements}
We would like to thank Antoine Petit, whose insightful comments improved the quality and clarity of this manuscript.
We are also grateful for additional comments and conversations with Eritas Yang, Mickey Teekamongkol, Lizzy Jones, Ivan Dudiak, Jessica Lin, Waldo Cardenas, Eric Agol and Eugene Chiang.
The presented numerical calculations were made possible by computational resources provided through an endowment by the Albrecht family.
\end{acknowledgements}

\appendix

\section{Disturbing Function Expansion Approach} \label{sec:disturbing}

The analogy to the dynamics of a pendulum in the traditional celestial mechanics approach is straightforward in the CR3BP.
The test particle is subject to a gravitational potential (per unit mass) $V_{int}$ from the massive planet that is periodic in the various orbital angles.
One can therefore Fourier expand $V_{int}$ in the orbits' mean longitudes $\lambda_i$; if one additionally expands the Fourier amplitudes as a power series in the test particle's orbital eccentricity $e_2$ (assumed small), this yields to leading order in the eccentricity \citep{Murray99},
\begin{eqnarray}
V_\text{int} &=&  -\frac{Gm_1}{a_{20}}
\sum_{jk}^{k}C_{jk}(\alpha)e_2^{k}\cos\left(\phi_{jk}\right) \nonumber \\
\phi_{jk} &=& j\lambda_2 - (j-k)\lambda_1-k\varpi_2, \label{Eres}
\end{eqnarray}
where $G$ is the gravitational constant, $m_1$ the inner planet's mass, $a_{20}$ the test particle's initial semimajor axis, and $\varpi_2$ its longitude of pericenter.
The $C_{jk}$ coefficients are treated as constant functions of the initial semimajor-axis ratio $\alpha \equiv a_{10}/a_{20}$.

The idea that allows for analytical progress is that most of the angle combinations $\phi_{jk}$ are cycling from 0 to $2\pi$ on a timescale equal to or shorter than the orbital period, so that on the long timescales over which the orbits evolve, these harmonics simply average out.
This suggests searching for combinations of angles $\phi_{jk}$ that vary slowly, and whose effects can build up in phase over many orbits and yield larger-amplitude variations. 
For example, when planets have mean motions $n_i \equiv \dot{\lambda}_i$ (with overdots denoting time derivatives) that are near an integer ratio such that $pn_2 \approx (p-q)n_1$, the combination $\phi_{pq}$ varies slowly, since the mean motions are much faster than the pericenter precession rates, and thus $\dot{\phi} \approx pn_{20} - (p-q)n_{10} \approx 0$.
Therefore, if one treats $e_2$ as a constant parameter and isolates only the resonant harmonic, one obtains the potential of a simple pendulum $V_{int} \propto \cos{\phi}$. 

\section{The Relative and Center-of-Mass Eccentricity in the Hill Problem} \label{sec:erelHill}

In the case of two massive planets on eccentric orbits, one needs to be a bit more careful.
First, because the inner planet no longer orbits at a uniform rate, we choose to work in a frame rotating uniformly at the mean motion $n$ corresponding to the ``center-of-mass" semimajor axis $a$ given by Eq.\:\ref{eq:a}.

Second, with two independent orbits, we need to fix things to a common coordinate system.
In particular, we define $t=0$ as the time when both orbits' guiding centers cross the inertial direction from which the pericenters $\varpi_i$ are measured.
Because we assume the eccentricities and deviations $\Delta a_i$ from the center-of-mass semimajor axis are already small, we approximate $a_1 \approx a_2 \approx a$ and $n_1 \approx n_2 \approx n$ in terms that already involve the $e_i$ and $\Delta a_i$, and ignore the additional corrections quadratic in these small quantities.
This yields,
\begin{eqnarray}
x_1 &=& ae_1 \cos(nt-\varpi_1) + \Delta a_1 \nonumber \\
x_2 &=& ae_2 \cos(nt-\varpi_2) + \Delta a_2 \nonumber \\
y_1 &=& -2ae_1 \sin(nt - \varpi_1) - \frac{3}{2} \Delta a_1 nt \nonumber \\
y_2 &=& -2ae_2 \sin(nt - \varpi_2) - \frac{3}{2} \Delta a_2 nt \label{eq:xi}
\end{eqnarray}

If we now calculate the relative separations and expand the sines and cosines,
\begin{eqnarray}
x_r &=& x_2 - x_1 = a\Bigg[(e_2 \cos \varpi_2 - e_1 \cos \varpi_1) \cos(nt) + (e_2 \sin \varpi_2 - e_1 \sin \varpi_1) \sin(nt) \Bigg] + \Bigg(\Delta a_2 - \Delta a_1 \Bigg) \nonumber \\
y_r &=& y_2 - y_1 = -2a\Bigg[(e_2 \cos \varpi_2 - e_1 \cos \varpi_1) \sin(nt) - (e_2 \sin \varpi_2 - e_1 \sin \varpi_1) \cos(nt) \Bigg] - \frac{3}{2} \Bigg(\Delta a_2 - \Delta a_1 \Bigg) nt \label{eq:xrlong}
\end{eqnarray}

Defining
\begin{equation}
{\bf e_r} \equiv {\bf e_2} - {\bf e_1} \label{eq:er}
\end{equation}
so that
\begin{eqnarray}
e_{rx} \equiv e_r \cos{\varpi_r} &=& e_2 \cos \varpi_2 - e_1 \cos \varpi_1 \\
e_{ry} \equiv e_r \sin{\varpi_r} &=& e_2 \sin \varpi_2 - e_1 \sin \varpi_1 \\
\end{eqnarray}
we can recombine Eqs.\:\ref{eq:xrlong} to obtain
\begin{eqnarray}
x_r &=& ae_r \cos(nt-\varpi_r) + \Delta a \nonumber \\
y_r &=& -2ae_r \sin(nt - \varpi_r) - \frac{3}{2} \Delta a nt \\
\end{eqnarray}
where $\Delta a = \Delta a_2 - \Delta a_1 = a_2 - a_1$ is the total semimajor axis separation.

This is the epicyclic form for a single Keplerian orbit (cf. Eq.\:\ref{eq:xi}) with semimajor axis offset from the center-of-mass semimajor axis $a$ by $\Delta a$, with eccentricity vector ${\bf e_r}$. 

Similarly, we have

\begin{eqnarray}
x_{COM} &=& \frac{m_1 x_1 + m_2 x_2}{m_1 + m_2} = a\Bigg[\frac{m_1 e_1 \cos \varpi_1 + m_2 e_2 \cos \varpi_2}{m_1+m_2} \cos(nt) + \frac{m_1 e_1 \sin \varpi_1 + m_2 e_2 \sin \varpi_2}{m_1+m_2} \sin(nt) \Bigg] + \frac{m_1\Delta a_1 + m_2 \Delta a_2}{m_1 + m_2} \nonumber \\
y_{COM} &=& \frac{m_1 y_1 + m_2 y_2}{m_1 + m_2} = -2a\Bigg[\frac{m_1 e_1 \cos \varpi_1 + m_2 e_2 \cos \varpi_2}{m_1+m_2} \sin(nt) - \frac{m_1 e_1 \sin \varpi_1 + m_2 e_2 \sin \varpi_2}{m_1+m_2} \cos(nt) \Bigg] - \frac{3}{2} \frac{m_1\Delta a_1 + m_2 \Delta a_2}{m_1 + m_2} nt, \label{eq:xcomlong}
\end{eqnarray}
where the final terms in both expressions are zero since
\begin{equation}
\frac{m_1\Delta a_1 + m_2 \Delta a_2}{m_1 + m_2} = \frac{m_1 a_1 + m_2 a_2}{m_1 + m_2} - a = 0,
\end{equation}
using Eq.\:\ref{eq:a}. 

Defining
\begin{equation}
{\bf e_{COM}} \equiv \frac{m_1{\bf e_1} + m_2{\bf e_2}}{m_1 + m_2}
\end{equation}
so that
\begin{eqnarray}
e_{COMx} \equiv e_{COM} \cos{\varpi_{COM}} &=& \frac{m_1 e_1 \cos \varpi_1 + m_2 e_2 \cos \varpi_2}{m_1+m_2} \\
e_{COMy} \equiv e_{COM} \sin{\varpi_{COM}} &=& \frac{m_1 e_1 \sin \varpi_1 + m_2 e_2 \sin \varpi_2}{m_1+m_2} \\
\end{eqnarray}
we can recombine Eqs.\:\ref{eq:xcomlong} to obtain
\begin{eqnarray}
x_{COM} &=& ae_{COM} \cos(nt-\varpi_{COM}) \nonumber \\
y_{COM} &=& -2ae_{COM} \sin(nt - \varpi_{COM}) \\
\end{eqnarray}
which is the epicyclic form for a Keplerian orbit with semimajor axis $a$ and eccentricity $e_{COM}$.

Together with the mass-dependence discussed in Sec.\:\ref{sec:mass}, in the Hill limit we can therefore map the general case of two massive eccentric orbits to the circular restricted case, where one assigns all the mass to the inner planet on a circular orbit with the center-of-mass semimajor axis $a$ (Eq.\:\ref{eq:a}), and places an outer test particle at a separation $\Delta a$ with eccentricity vector ${\bf e_r}$ (Eq.\:\ref{eq:er}).
The pericenter of this relative orbit always points to the location where the two orbits come closest together.

\section{Corrections for Moderate Spacings} \label{sec:spacings}

Throughout this paper, we have worked in the Hill limit where the orbits are closely separated.
We now use the more general results from \cite{Hadden19} to explore the leading order corrections for moderate spacings, and the substantial corrections for the 2:1 MMR in particular.

\subsection{Mass Correction}
Starting from the disturbing function, \cite{Hadden19} shows that, in detail, the mass parameter should be defined as
\begin{equation}
\mu_\text{corr} = \frac{m_1 + \alpha m_2}{M_\star},
\end{equation}
where $\alpha = a_1/a_2 \approx 1$ for closely spaced orbits.
We see therefore that the corrected mass factor $\mu_\text{corr}$ can only be smaller than our simple mass factor $\mu$ above (Eq.\:\ref{eq:mu}).
If we define a fractional mass correction factor $m_\text{corr}$ so $\mu_\text{corr} = \mu (1-m_\text{corr})$, then using that $e_c \approx 1-\alpha$,
\begin{equation}
m_\text{corr} = \frac{m_2}{m_1+m_2} e_c.
\end{equation}
We see that the correction $m_\text{corr} < e_c$ and thus becomes negligible in the closely spaced limit as the crossing eccentricity approaches zero.
We also see that no correction is needed in the case where all the planetary mass in the inner planet, and that the worst case is when all the mass in the outer planet as $m_\text{corr} = e_c$.

Because the MMR widths and frequencies scale as $\mu^{1/2}$ (Sec.\:\ref{sec:widthsgen}), the correction to these quantities is $(1-m_\text{corr})^{1/2} \approx (1 - m_\text{corr}/2)$.
Thus, while the widths and frequencies are slightly smaller than our simple expressions suggest, these corrections are $\lesssim 10$\% even out as far as the $3:2$ MMR where $e_c = 2q/3p \approx 0.2$.

\subsection{Eccentricity Correction}
\cite{Hadden19} additionally shows that, in general, the relative eccentricity (cf. Eq.\:\ref{eq:erel}) should be defined as
\begin{equation}
{\bf e_{12}} = {\bf e_{2}} - \beta{\bf e_{1}}, \label{eq:erelcorr}
\end{equation}
where 
\begin{equation}
\beta \approx \Bigg(\frac{P_1}{P_2}\Bigg)^{0.55} \approx 1 - 0.83 e_c,
\end{equation}
so $\beta \approx 1$ and reduces to our simple expression for closely spaced orbits with small $e_c$.

In this case, there is no correction if all the eccentricity is in the outer planet (e.g., an eccentric test particle with an inner massive perturber). 
In the opposite limit where all the eccentricity is in the inner orbit, the relative eccentricity is smaller than our simple vector difference (Eq.\:\ref{eq:erel}) by $(1-0.83e_c)$.
This implies that the corresponding MMR widths and frequencies are smaller by $(1-0.83e_c)^{q/2} \approx (1-0.42qe_c)$.
For first-order MMRs ($q=1$) this corresponds again to a correction $\lesssim 10$\% as far out as the 3:2.
However, the correction for higher order MMRs at comparable separations is more significant.
For example, for the fifth order 14:9 MMR, putting all the eccentricity in the inner orbit results in an MMR width and frequency that is $\approx 40$\% smaller than putting all the eccentricity in the outer orbit and using Eq.\:\ref{eq:erel} for the relative eccentricity.

Finally, we caution that in cases where the dynamics don't behave as expected, one should check that neither the simple (Eq.\:\ref{eq:erel}) nor the corrected (Eq.\:\ref{eq:erelcorr}) relative eccentricity approach zero.
In the former case, the simple relative eccentricity (Eq.\:\ref{eq:erel}) can be several times too small compared to the corrected value (Eq.\:\ref{eq:erelcorr}).
In the latter case, one is not in the pendulum approximation even when using the corrected relative eccentricity, and the second fundamental model of resonance should be used (Sec.\:\ref{sec:limits}).

\section{The special case of the 2:1 MMR} \label{sec:21}

A planet can act on its neighbor either directly through the gravitational force between them, or by exerting a gravitational force on the star, which indirectly affects its neighbor by causing the star to accelerate.
The former give rise to so-called ``direct" terms in the disturbing function expansion, which include cosine terms corresponding to particular MMRs \citep{Murray99}, which we have obtained in this paper through an alternate, physically motivated approach.
The latter give rise to ``indirect" terms in the disturbing function.
These are rarely discussed because their effects typically average out so they do not qualitatively affect the dynamics \citep{Murray99}.

However, it is well known that there are indirect terms that have the same cosine arguments as $N:1$ MMRs \citep[for an explicit derivation, see][]{celmech}, so near a $N:1$ MMR, the corresponding indirect term does not average out, and significantly affects the dynamics.
None of the $N:1$ MMRs are really in the closely packed regime we explore in this paper.
However, the 2:1 MMR lies at fairly close separations, and because it is a strong (and therefore common) first-order MMR, we consider it for completeness.

Starting from the equations of \cite{Hadden19}, one can show after significant algebra, that the appropriate generalization of the relative eccentricity for the case of a 2:1 MMR is

\begin{equation}
{\bf e_{12}} = 0.27{\bf e_{2}} - 0.74{\bf e_{1}}. \label{eq:erel21}
\end{equation}

We note that this relative eccentricity no longer points toward the location where the orbits are closest together, and does not have a ready geometrical interpretation.
Indeed, we do not have a simple physical explanation for why only $N:1$ MMRs are modified in this way.

Equation \eqref{eq:erel21} also shows that the appropriate relative eccentricity is always smaller than the naive estimate (Eq.\:\ref{eq:erel}).
In the limit where all the eccentricity is in the inner planet, the width and frequency is only slightly shrunk to $0.74^{1/2} \approx 85$\% of the naive expectation that uses $e_{12} = e_1$.
However, in the opposite limit where all the eccentricity is in the outer planet, the width and frequency are cut by about half.

Finally, we point out that when going beyond the leading order in eccentricity explored in this paper, one considers additional harmonics, i.e. cosine terms with integer multiples of the resonant angle (whose coefficients involve higher powers of the eccentricity and are therefore typically small). 
These harmonics introduce quantitative corrections to the resonance width \citep[see][for discussion and a way to calculate these corrections numerically with \texttt{celmech}]{Hadden19}, which can become noticeable at high eccentricities (Fig.\:\ref{fig:widths}, see also \citealt{Wang17} for an alternate numerical investigation at high eccentricities).
Importantly, the dynamics still depend only on multiples of a single angle, and this therefore preserves an integrable one degree-of-freedom problem \citep[see Sec. 2.3 of][for a discussion]{Hadden19}.
For the 2:1 MMR, however, the indirect terms only affect the leading-order harmonic (modifying the eccentricity coefficients according to Eq.\:\ref{eq:erel21}), while the higher-order harmonics remain unchanged. 
This introduces two separate angles into the dynamics, rendering it a non-integrable 2 degree-of-freedom problem and opens up the possibility of chaos, particularly at the higher eccentricities where the higher order harmonics become appreciable.

\bibliography{Bib}

\bibliographystyle{aasjournal}
\end{document}